\documentclass[twocolumn,aps,prl,showpacs,amsmath,superscriptaddress]{revtex4-1}
\usepackage{amsfonts}
\usepackage{amssymb}
\usepackage{mathrsfs}
\usepackage{graphicx}
\usepackage{float}
\usepackage{xcolor}

\begin{document}

\title{Multi-particle quantum walks and Fisher information in one-dimensional lattices }

\author{Xiaoming Cai}
\address{State Key Laboratory of Magnetic Resonance and Atomic and Molecular Physics, Wuhan Institute of Physics and Mathematics, APM, Chinese Academy of Sciences, Wuhan 430071, China}
\author{Hongting Yang}
\address{School of Science, Wuhan University of Technology, Wuhan 430071, China}

\author{Hai-Long Shi}
\address{State Key Laboratory of Magnetic Resonance and Atomic and Molecular Physics, Wuhan Institute of Physics and Mathematics, APM, Chinese Academy of Sciences, Wuhan 430071, China}
\affiliation{University of Chinese Academy of Sciences, Beijing 100049, China.}

\author{Chaohong Lee}
	\email{lichaoh2@mail.sysu.edu.cn}
	\affiliation{Guangdong Provincial Key Laboratory of Quantum Metrology and Sensing $\&$ School of Physics and Astronomy, Sun Yat-Sen University (Zhuhai Campus), Zhuhai 519082, China}
	\affiliation{State Key Laboratory of Optoelectronic Materials and Technologies, Sun Yat-Sen University (Guangzhou Campus), Guangzhou 510275, China}

\author{Natan Andrei}
\affiliation{Department of Physics, Rutgers University, Piscataway, New Jersey 08854, USA}

\author{Xi-Wen Guan}
\email{xwe105@physics.anu.edu.au}
\affiliation{State Key Laboratory of Magnetic Resonance and Atomic and Molecular Physics,
Wuhan Institute of Physics and Mathematics, APM, Chinese Academy of Sciences, Wuhan 430071, China}
\affiliation{NSFC-SPTP Peng Huanwu Center for Fundamental Theory, Xi'an 710127, China}
\affiliation{Department of Theoretical Physics, Research School of Physics and Engineering,
Australian National University, Canberra ACT 0200, Australia}

\date{\today}

\begin{abstract}

 Recent experiments on quantum walks (QWs)   demonstrated a full control over the statistics-dependent walks of single and two particles in one-dimensional (1D) lattices.
However little is known about the general characterization  of QWs   at the many-body level.
Here  we rigorously  study QWs, Bloch oscillations and quantum Fisher informations (FIs) for three indistinguishable bosons and fermions   in 1D lattices using time-evolving block decimation algorithm and many-body perturbation theory.
We show that such strongly correlated  QWs  not only give rise to statistics-and-interaction-dependent ballistic transports of scattering states and of two- and three-body bound states,
but also  allow  a quantum enhanced precision measurement of the gravitational force.
In contrast to the QWs of the fermions, the QWs of three  bosons exhibit strongly correlated Bloch oscillations, which  present a surprising time scaling $t^3$ of FI  below a characteristic time $t_0$ and saturate to  the fundamental limit of $t^2$ for $t>t_0$.

\end{abstract}
\maketitle

Quantum walks (QW) \cite{Aharonov1,Kempe1}, the quantum counterpart of the classical random walks,  are characterized by a fast ballistic spreading with wave fronts expanding linearly  in  time.
Owing to their non-classical features, they  have potential  applications in quantum algorithms \cite{Ambainis1}, quantum computing \cite{Childs1}, quantum information \cite{Andraca1,Zatelli:2020}, quantum simulation \cite{Asboth1} and quantum biology \cite{Lloyd1}.
QWs have been experimentally implemented in a variety of quantum systems \cite{Wang1}, and recently found in detecting  topological states \cite{Kitagawa1,Kraus1,Ramasesh1}, discrete-time QWs \cite{Cedzich1,Arnault1,Destri1,Bisio1},   and bound states of magnons \cite{Ahlbrecht1,Fukuhara1}.

Up to now,  most previous works focused on the one and two-particle QWs in 1D lattices.
Preliminary experiments studied the QWs of single and two particles by using either neutral atoms \cite{Karski1}, ions \cite{Schmitz1}, photons \cite{Schreiber1}, spin impurities \cite{Fukuhara1,Fukuhara2}, or nuclear-magnetic-resonance systems \cite{Du1}.
Walkers of two non-interacting particles can develop non-trivial correlations due to the Hanbury-Brown-Twiss  interference \cite{Peruzzo1,Mayer1,Hillery1,Sansoni1,Solntsev1,Lahini1,Brown1}.
Bosonic (fermionic) walkers result in an emergence of bunching (anti-bunching) in density-density correlations \cite{Omar1,Benedetti1,Qin1,Preiss1}, and anyons are in between \cite{Wang2,Lau:2020}.
Moreover,  the interplay between quantum statistics and interaction of two particles \cite{Iyer:2013,Preiss1,Ganahl1,Krapivsky1,Wiater1,Qin1,Siloi:2017,Beggi:2018,Sarkar:2020}, and of two and three flipped spins   in a Heisenberg Chain \cite{Liu1}, leads to a richer dynamics of quantum co-walking.

On the other hand, a quantum particle in a tilted periodic potential may undergo Bloch oscillation (BO), which has been demonstrated via ultracold atoms \cite{Preiss1,Geiger1}.
The BO frequency is proportional to the tilting force.
%
This can be employed to measure the gravitational force \cite{Ferrari1,Tarallo1},  the magnetic field gradient \cite{LiuWJ}, the Zak phase in topological Bloch bands \cite{Atala1} and the Casimir-Polder force \cite{Carusotto1}.
Quantum Fisher Information (FI), which provides a lower limit to the Cram\'{e}r-Rao bound, plays a central role in  quantum precision measurements \cite{Helstrom:1976,Pezze1,Liu2}.
However, the question how to create many-body entanglement to improve measurements precision via BOs  and how to use the FIs to quantify the precision limit for the gravitational force  still remains open and challenging.

In this Letter,  we study nonequilibrium dynamics of three fermions and bosons in 1D lattices and explore its metrological application in precision measurement of gravitational force.
Continuous-time QWs,  strongly correlated BOs, the band structure, time-evolutions of density distributions and density-density correlations for the systems are thoroughly studied through both numerical and analytical methods.
%
%
The FIs of such QWs  show  a promising capability of quantum-enhanced  precision measurement of weak forces in the walks of three-boson bound states.

\emph{The Model.--} We consider three indistinguishable particles moving on 1D lattices  governed by the Hamiltonian:
\begin{equation}
\hat{H}=-J\sum_{j=-M}^{M-1}(a^\dagger_ja_{j+1}+\mathrm{H.c})+V\sum_{j=-M}^{M-1}\hat{n}_j\hat{n}_{j+1}.
\label{EQN1}
\end{equation}
Here $a^\dagger_j$ ($a_j$) creates (annihilates) a particle at the $j$-th site, and $\hat{n}_j=a^\dagger_ja_j$ is the particle number operator.
The total number of lattice sites $L=2M$.
$J$ is the nearest-neighbour hopping and set  the unit of energy ($J=1$).
$V$ is the nearest-neighbour interaction.
Hamiltonians of this kind (\ref{EQN1}) can be realized with ultracold atoms \cite{Preiss1,Sun:2020e,Yang:2020e}.
Here we study two  types of particles: bosons and fermions.
The fermonic model is  equivalent to the exactly solvable XXZ Heisenberg chain \cite{YY:1966,Takahashi1} which is also equivalent to the hard-core bosonic model \cite{Jordan1}.  
The QW of two and three particles  of the XXZ model  was found in \cite{Liu1} and  an experimental realization 
   via ultracold two-level atoms in deep optical lattices was  given in \cite{Lee2004,Qin1,LiuWJ}. 

%
%

%

\begin{figure}[tbp]
\begin{center}
\includegraphics[scale=0.65, bb=46 53 416 325]{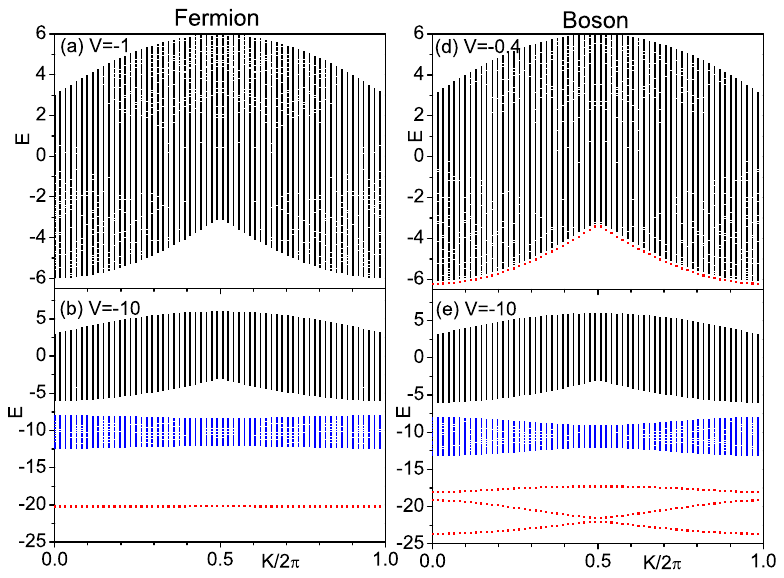}
\includegraphics[scale=0.65, bb=43 31 418 183]{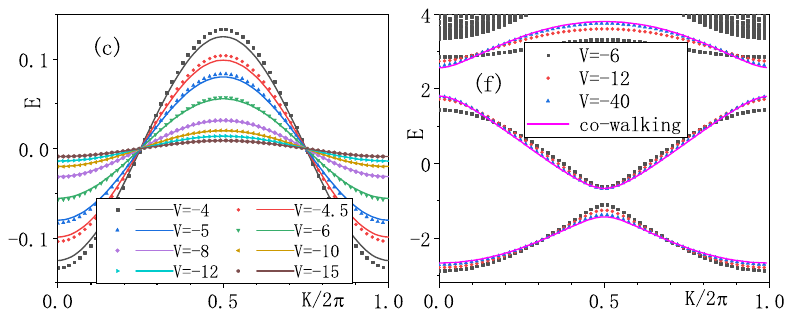}
\caption{(Color online) The spectra of three fermions (a,\,b)  and bosons (d,\,e)  for $L=61$ and different values of interaction strength.
Each point represents an eigenenergy $E$ for a given total momentum $K$, and the red and blue ones correspond to bound states.
(c) and (f) show the spectra of 3BSs of fermions  and bosons, respectively. The solid lines denote the perturbation results  of  (\ref{EFF1}) and (\ref{EFF2}) which agree well with  the ED calculation (symbols).
All spectra in either  (c) or (f) are shifted  by a constant.
}
\label{Fig1}
\end{center}
\end{figure}
\emph{Spectra and quantum walks.--}

%
Within the three-particle Hilbert space,
we first perform  exact diagonalization (ED) of the systems  in momentum space.
Fig.\ref{Fig1} shows   spectra of three fermions and bosons,  respectively.
We observe that the three-particle systems of bosons and fermions  host scattering states (SSs), two-body bound states (2BSs) and three-body bound states (3BSs).
In the weak  interaction region,  the spectra only contain one continuum band corresponding to SSs for both three bosons and fermions.
However, as the attraction increases the spectrum behave rather statistics-dependently, see Fig.\ref{Fig1}(a,b,d,e).
For the bosonic system, the  BSs split from the continuum band  when the interaction $V$  becomes stronger.
The whole spectra contain three isolated spectra  with gaps in between.
Three mini-bands of the  3BSs,  which are  energetically lower than that of the 2BSs (blue part in Fig.\ref{Fig1}(b,e)), remarkably form competitive QWs in time evolution.
The SSs band  is continuous with highest energies.
%
%
In contrast, for the fermionic system there is only one continuum SSs  band as long as the interaction $|V|<1$.
%
%
Bands of the  SSs, 2BSs and 3BSs are  energetically separated for  large attractions.
The 3BSs only constitute one mini-band with the lowest  energy, and it becomes more and more flat when the attraction increases.

Now we employ time evolving block decimation (TEBD) algorithm \cite{Vidal1,Iyer:2013} to numerically  simulate the three-particle continuous-time QWs. 
They are governed by the unitary time-evolution $|\psi(t)\rangle=e^{-iHt}|\psi(t=0)\rangle$, in contrast to the discrete-time QWs which obey a successive single-time evolution governed by `shift' and `coin' operators \cite{Cedzich1,Arnault1,Destri1,Bisio1}.
Here we set the initial state $|\psi(t=0)\rangle=a^\dagger_{-1}a^\dagger_{0}a^\dagger_{1}|0\rangle$ with three particles at the three central neighbouring sites.
Open boundary conditions are used under the  cumulative truncation errors in the order of $10^{-8}$.
We study time evolutions of density distribution $n_j(t)=\langle\psi(t)|a^\dagger_ja_j|\psi(t)\rangle$ and density-density correlation function $C_{i,j}(t)=\langle\psi(t)|\hat{n}_i\hat{n}_j|\psi(t)\rangle$, which preserve the symmetries $n_j(t)|_V=n_j(t)|_{-V}$ and $C_{i,j}(t)|_V=C_{i,j}(t)|_{-V}$ \cite{Yu1,Supplemental}. %
We thus   only  consider the attractive interaction  in our study.

\begin{figure}[tbp]
\begin{center}
\includegraphics[scale=0.67, bb=144 223 477 629]{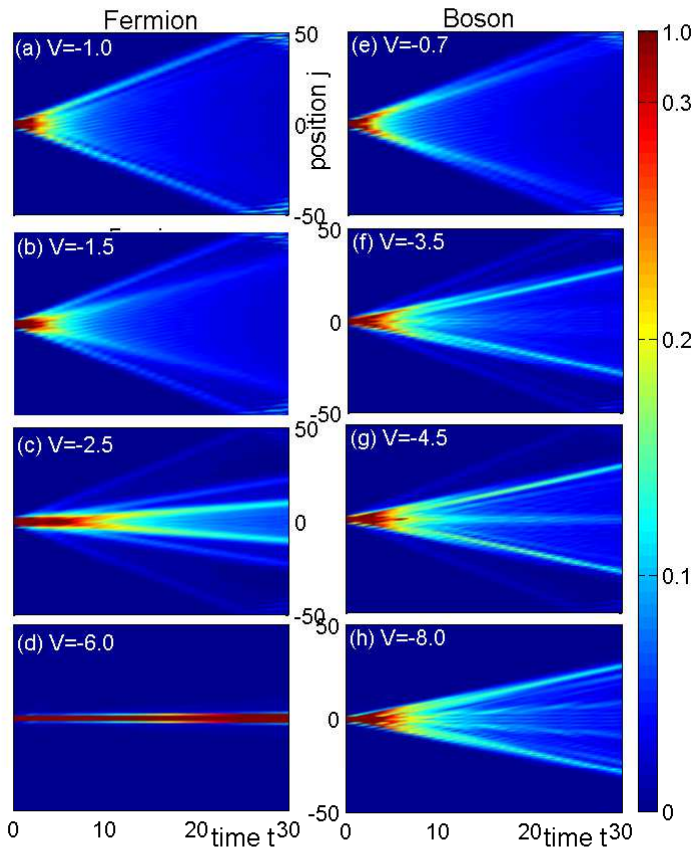}
\caption{(Color online) Time-evolutions of density distribution for three-fermion (a-d) and boson (e-h) systems with $L=101$ and different values of attractions, see text.
Results   (a)-(d)    match  analytic results for the XXZ Heisenberg model \cite{Liu1}.
}
\label{Fig2}
\end{center}
\end{figure}

In Fig.\ref{Fig2} we show time-evolutions of density distribution for three-particle QWs.
Both the weakly interacting bosons  and fermionic systems with $|V|<1$ have the same form of evolution, i.e.
a single  ballistic expansion light-cone is established because of the single  continuum band of  the  SSs.
For the fermionic system with the interaction $|V|>1$,  an inner cone emerges with a slower and linearly moving wave front, indicating  the formation of BSs.
Continuing to increase the interaction,  the third   innermost cone forms.
From outer  to inner,  three cones correspond to ballistic expansions of the SSs, 2BSs and 3BSs, respectively, see Fig.\ref{Fig2}(c).
As $|V|$ is further increased, the cones of  SSs and 2BSs gradually fade away  and only the light-cone  of  the  3BSs remains.
The speed of wave front (SWF) of the  SSs is  independent of interaction,  i.e., it  is always $2$, showing a maximal group velocity (MGV) of non-interacting particles.
But SWFs for 2BSs and 3BSs decrease when the interaction $|V|$ increases.  Note that the results of Fig.  2  (a)-(d)  for fermions  match nicely  the corresponding analytic results
for the equivalent  XXZ Heisenberg model \cite{Liu1}.

For the  bosonic system, besides the outer cone for SSs,  an inner cone for the BSs emerges as long as $V$ is nonzero.
When the interaction increases, the SWF of this inner cone first decreases and then stops decreasing at a fixed value due to the band structure of the 3BSs.
Then the third innermost cone shows up, its SWF first decreases and then  increases to a finite value, see Fig.\ref{Fig2} (f), (g), (h) and Fig.S9 in  \cite{Supplemental}.
%
%
%
This unique behaviour of innermost cone is caused by the interplay of  the 2BS and the 3BS.
%
%
For a large enough attraction, the evolution contains only two cones which are both related to the three mini-bands of 3BSs.
%
%

\begin{figure}[tbp]
\begin{center}
\includegraphics[scale=0.8, bb=57 313 313 545]{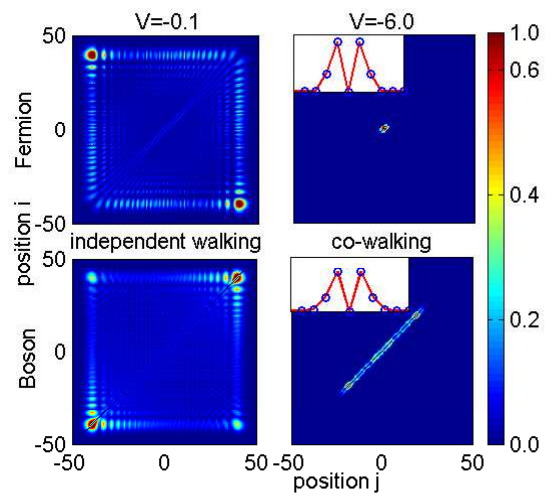}
\caption{(Color online) Density-density correlation functions $C_{i,j}$ for both fermionic (upper panel) and bosonic (lower panel) systems with a size  $L=101$ at the time $t=22$. Corresponding profiles $C_{i=0,j}$ around $j=0$ are shown in subsets.
}
\label{Fig3}
\end{center}
\end{figure}

Moreover, the density-density correlation function $C_{i,j}(t)$ also  provides an important  statistical nature  of  the three-particle QWs \cite{Supplemental}.
It significantly marks  the difference between  co-walking  and individual  walking.
The  co-walking particles bound together and move as a single composite particle, revealing the togetherness of quasiparticles.
The density-density correlations for co-walking show few  lines (5 lines)  at ($i=j\pm d$) with $d=1,\,2$  in the ($i,j$) plane, a signature of the co-walking.
In Fig.\ref{Fig3}, we show $C_{i,j}(t)$ for both fermionic and bosonic systems at time $t=22$ (they are free from the boundary effects).
For  a small $|V|$,  the  correlation function shows (anti-)bunching behaviour with (off-)diagonal correlations at the wave front in the (fermionic) bosonic system.
As $|V|$ increases, bunching and anti-bunching correlations fade away, and correlations on four minor diagonal lines ($i=j\pm1,\,2$) are gradually enhanced with respect to a statistics-dependent pace, see the subsets in Fig.\ref{Fig3}.
In contrast to the co-walking of two bosons \cite{Qin1}, the co-walking of three bosons remarkably shows expansive wave fronts due to the existence of the mini-bands of the 3BSs.

\emph{Many-body perturbation  and Bloch oscillations.--} Under a strong attraction, one can treat the hopping as a perturbation to the interaction term in the Hamiltonian (\ref{EQN1}), see \cite{Takahashi1}.
After projecting onto the subspace of the 3BSs, an effective single-particle model can be derived explicitly.
For the fermionic case, by using the  third order perturbation, an  effective single-particle Hamiltonian for the  co-walking of three fermions is given by  \cite{Supplemental}
\begin{equation}
\hat{H}^\mathrm{F}_{\mathrm{eff}}=-\frac{J^3}{V^2}\sum_j(c^\dagger_jc_{j+1}+\mathrm{H.c.}),
\label{EFF1}
\end{equation}
where  $c^\dagger_j=a^\dagger_{j-1}a^\dagger_ja^\dagger_{j+1}$.
The spectrum  of  this single-particle Hamiltonian (\ref{EFF1}) reads $E^\mathrm{F}_{\mathrm{eff}}(K)=-\frac{2J^3}{V^2}\mathrm{cos}(K)$  with a MGV $v^\mathrm{F}=2J^3/V^2$, see \cite{footnote2}. 
In Fig.\ref{Fig1}(c), we show the spectrum of the 3BSs for the fermionic system,  where the dots denote numerical result obtained  from ED and lines are obtained from the effective single-particle Hamiltonian (\ref{EFF1}).
Both results agree well as $|V|$ increases.
The MGV $v^\mathrm{F}$ is also in a good agreement with the SWF of the 3BSs \cite{Supplemental}.
We observe from Eq.(\ref{EFF1})  that the ballistic expansion of  three-fermion co-walking  becomes very slow as $|V|$ increases.

\begin{figure*}[tbp]
\begin{center}
\includegraphics[width=\linewidth, bb=4 339 579 532]{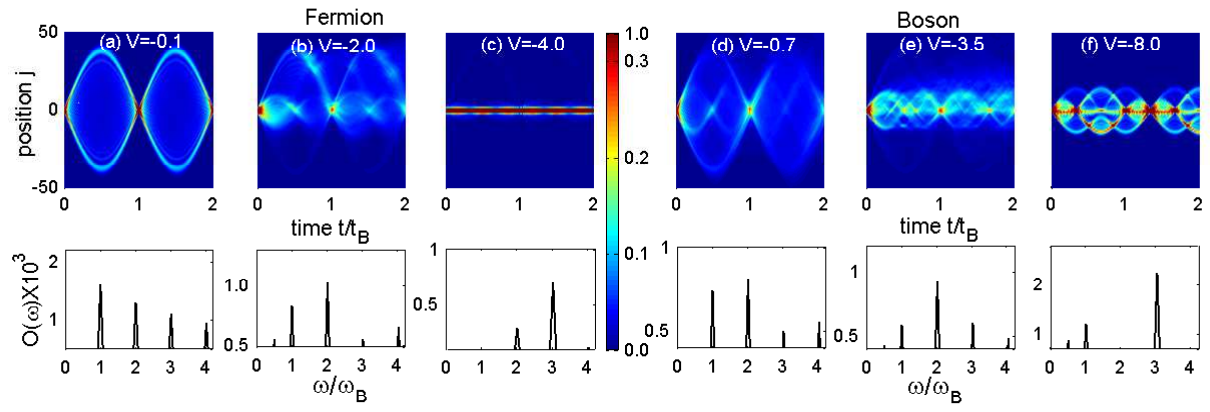}
\caption{(Color online) Left (a-c) and right (d-f) panels show  the time evolutions of density distributions for three and fermions with $L=101$ and $F=0.1$, respectively.  Multiple fractional Bloch oscillations are observed.  The corresponding  frequencies are shown in the bottom row.
 }
\label{Fig4}
\end{center}
\end{figure*}

The subspace of  three-boson co-walking has $3L$-fold degeneracies.
The first order perturbation gives an effective single-particle Hamiltonian \cite{Supplemental}
\begin{equation}
\hat{H}^\mathrm{B}_\mathrm{eff}=-\sqrt{2}J\sum_j(d^\dagger_jb_j+c^\dagger_jb_j+\sqrt{2}d^\dagger_{j+1}c_j+\mathrm{H.c.})
\label{EFF2}
\end{equation}
with three species $b^\dagger_j=a^\dagger_{j-1}a^\dagger_{j}a^\dagger_{j+1}$, $c^\dagger_j=\frac{1}{\sqrt{2}}(a^\dagger_{j})^2a^\dagger_{j+1}$, and $d^\dagger_j=\frac{1}{\sqrt{2}}a^\dagger_{j-1}(a^\dagger_{j})^2$.
Obviously,  it is independent of $V$.
In the momentum space we can get the spectra of the effective Hamiltonian (\ref{EFF2}), which are shown in Fig.~\ref{Fig1} (f) (solid lines), agree well  with the ED numerical results (doted lines).
The 3BSs have  three mini-bands which show  two different MGVs, $v^\mathrm{B}_1\simeq1$ for the middle mini-band and $v^\mathrm{B}_2\simeq0.64$ for other two mini-bands in Fig.~\ref{Fig1} (f) (solid lines).
$v^\mathrm{B}_{1(2)}$ agrees well with the SWF of the outer  (inner) cone when  $|V|\gg1$, see \cite{Supplemental}.

In order to achieve   a metrological application of QWs, we  add  a static force to the Hamiltonian (\ref{EQN1})
 \begin{equation}
\hat{H}_\mathrm{Force}=F\sum_{j}ja^\dagger_ja_j
\end{equation}
 with $F$ the strength of the applied force  and consider the  BOs from the same initial state $|\psi(t=0)\rangle=a^\dagger_{-1}a^\dagger_{0}a^\dagger_{1}|0\rangle$.

We illustrate the  BOs for fermionic systems in the left panel of Fig.\ref{Fig4}.
For a weak interaction, i.e. $|V|<1$, particles independently undergo a single-particle BO with the amplitude $4J/F$ and the temporal period $t_B=2\pi/F$ (frequency $\omega_B=F$) \cite{Kosevich1,Zhang1}.
When  $|V|>1$,  two inner BOs appear successively with smaller amplitudes and shorter periods.
From outer to inner  there are BOs of the SSs, 2BSs and 3BSs, respectively.
Upon further  increasing  the attraction, the two outer  BOs gradually fade away and only the BO of the  3BSs remains, see \cite{Supplemental}.
In order to analyze the periodicity, we introduce a density difference  $O(t)=\sum_{j}|n_j(t)-n_j(t=0)|/L$.
From the Fourier transformation $O(\omega)$, we observe the characteristics periodicities  of  $t_\mathrm{B}/2$ and $t_\mathrm{B}/3$ (or frequencies $2\omega_B$ and $3\omega_B$) BOs for 2BSs  and 3BSs, respectively,  which are called fractional BOs  in interacting systems \cite{Dias1,Khomeriki1,Corrielli1}.
$O(\omega)$ presents the relative weights  of BO
modes with different frequencies.
%
%
In the  strong coupling  limit, i.e.  $|V|\gg1$, the co-BO of  3BSs of fermions can be  described by the  effective single-particle Hamiltonian $\hat{H}^F_\mathrm{BO}=\hat{H}^\mathrm{F}_{\mathrm{eff}}+3F\sum_jjc^\dagger_jc_j$ with the BO amplitude  $4J^3/(3V^2F)$, which is  inversely proportional to $V^2$.
Here the effective force is $3F$, which leads to  the periodicity  of co-BO $t_\mathrm{B}/3$, as well as the frequency of $3\omega_\mathrm{B}$.

Due to the quantum  statistical difference, the ground-state degeneracies are different for bosons and fermions,  leading to different many-body perturbation processes as well as  different dynamics of the  BOs,  see  Fig.~\ref{Fig4}.
The corresponding  effective single-particle Hamiltonian, describing the BOs among  three mini-bands  under an effective force $3F$,  is given by
\begin{equation}
\hat{H}^B_\mathrm{BO}=\hat{H}^\mathrm{B}_{\mathrm{eff}}+3F\sum_j[jb^\dagger_jb_j+(j+\frac{1}{3})c^\dagger_jc_j+(j-\frac{1}{3})d^\dagger_jd_j],
\end{equation}
see \cite{Supplemental}.
We observe that the Landau-Zener tunellings display between two nearby mini-bands  \cite{Breid1,Longhi1}.
The amplitude of co-BO $\propto J/3F$  is independent of $V$, showing a larger FI than that of the co-walking of three fermions in next section.
Consequently, the periodicity  of co-BO of three bosons is $t_B/3$ and the  frequency is $3\omega_B$ that provides an ideal metrological state for a precision measurement of a weak force.

\begin{figure}[tbp]
\begin{center}
\includegraphics[scale=0.4, bb=0 0 454 354]{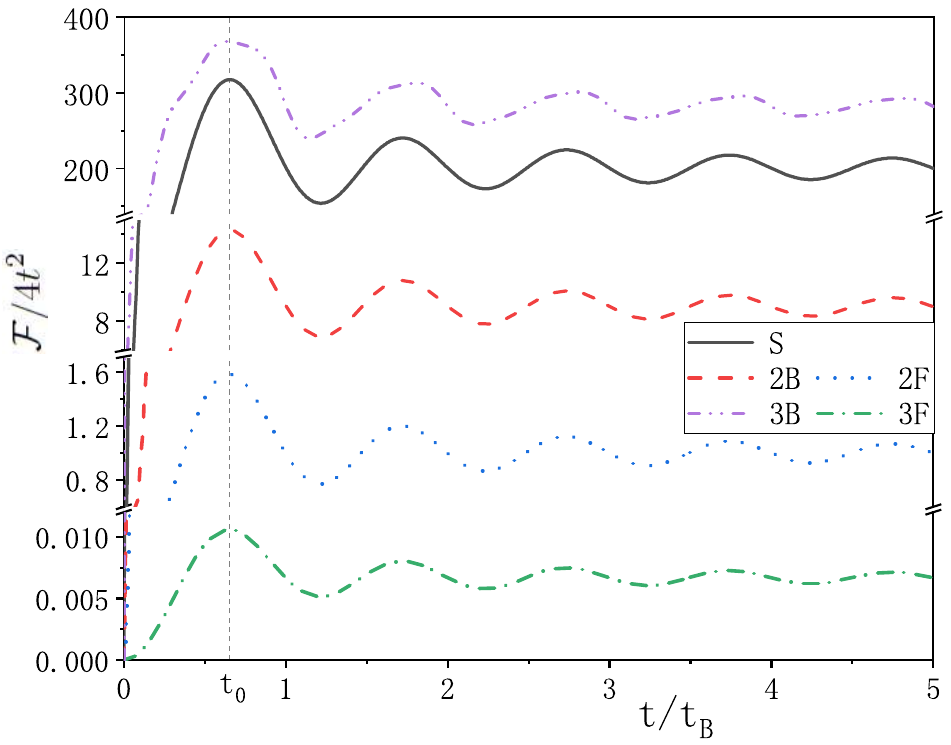}
\caption{(Color online) The time-dependent  quantum FIs  $\mathcal{F}/4t^2$ vs time $t$ for single- and multi-particle (co)-BOs. Parameters: L=101, V=-10, and $F=0.1$. The FIs are  proportional to $t^3$ for $t<t_0$, while $\mathcal{F} \sim t^2$ for $t>t_0$.  }
\label{Fig5}
\end{center}
\end{figure}

\emph{Fisher information and precision measurement.--}
%
%
 The three-boson QWs have a very rich dynamical structure  of the co-BOs   that leads to an almost interaction-independent co-BO amplitude and high value of FI in the probe of the weak force, see \cite{Supplemental}.
Here the FIs for (co-)BOs presents  the precision limit for single parameter $F$.
By definition of FI for an unitary process from a pure initial state  \cite{Braunstein1,Braunstein2,Liu2}, we can calculate the FIs of single- and multi-particle (co-)BOs \cite{Supplemental}
\begin{eqnarray}
\mathcal{F}&=&4[(\frac{\partial}{\partial F}\langle\psi(t)|)\frac{\partial}{\partial F}|\psi(t)\rangle-|\langle\psi(t)|\frac{\partial}{\partial F}|\psi(t)\rangle|^2],\notag\\
&=&4t^2\Delta H^Q(t),
\label{FI}
\end{eqnarray}
where $\Delta H^Q(t)$ is the fluctuation of a time-dependent effective Hamiltonian $H^Q(t)$ over the initial state $|\psi_0\rangle$.
\begin{eqnarray}
\Delta H^Q(t)&=&\langle\psi_0|[H^Q(t)]^2|\psi_0\rangle-\langle\psi_0|H^Q(t)|\psi_0\rangle^2,\notag\\
H^Q(t)&=&h[it\cdot\mathrm{ad}_{H_{\mathrm{BO}}}](\partial_FH_{\mathrm{BO}}),
\end{eqnarray}
with operator function $h[x]=(e^x-1)/x$ and  adjoint operator $\mathrm{ad}_G(C)=[G,C]$.
In Fig.\ref{Fig5} we show FIs  as functions of time for (co)-BOs.
Here we denoted single-particle (S), two-boson (2B), two-fermion (2F), three-fermion (3F), and three-boson (3B) (co-)BOs, respectively.
We demonstrate  that below a characteristic time $t_0\approx 0.5 t_B$, the  time scalings of FIs show a surprising  power law form $\mathcal{F}\simeq \alpha t^3$, where $\alpha$ is a state-dependent constant \cite{footnote},  also see \cite{Supplemental}.
 In contrast to the smallest FI  of the three-fermion co-BO with $\alpha \approx 0.024$, the three-boson co-BO has the largest FI  with the  largest value of $\alpha \approx 1185.485$,   then the smallest uncertainty in the measurement of force.
The single-particle BO has the second largest FI.
For $t>t_0$, the FIs for these walk states saturate to   the standard quantum limit $\mathcal{F}\simeq4At^2$ with case-dependent constant coefficients $A$.
%
%
%
%
%
%


\emph{Conclusions and Discussions.--} We have studied  continuous-time QWs, BOs and FIs of three bosons and fermions   in 1D lattices,  which   reveal   intrinsic and extrinsic  roles of quantum statistics,  interaction and  the gravitational force in the quantum random walks.
We have demonstrated  that the metrological useful entanglement for  high precision measurements of weak forces can be generated under the  time evolutions of suitable quantum states.
Our method also  holds a promise for a quantum-enhanced precision test of the EP through the  QWs of three bosons.

The superiority of three-boson co-BO in precision measurement of weak force would provide a potential approach to test the Einstein equivalence principle (EP).
Instead of making comparison with the  BOs of non-interacting isotopes \cite{Tarallo1}, one may  test the EP  through the BOs of same species of three interacting bosons, see a discussion in  \cite{Supplemental}.

This work is supported by the National Key R\&D Program of China  No.\ 2017YFA0304500 and the NKRDP under Grant No. 2016YFA0301503,   the NSFC grant No.\ 11874393, No.\ 1167420  and No.\ 12025509.
XMC, HTY and HLS equally contribute the numerical and analytical studies for this research.
The authors thank Jing Liu and Wei-Dong Li for helpful discussions.

\clearpage\newpage
\setcounter{figure}{0}
\setcounter{table}{0}
\setcounter{equation}{0}
\def\thefigure{S\arabic{figure}}
\def\thetable{S\arabic{table}}
\def\theequation{S\arabic{equation}}
\setcounter{page}{1}
\pagestyle{plain}

\section*{Supplementary material: Multi-particle quantum walks and Fisher information in one-dimensional lattices }

\section{S1. Symmetries in  dynamic evolution}

For the system with only nearest-neighbour interactions, we can decompose it into odd and even lattices, and define a symmetry operator $W$ which acts on the bipartite lattice via  \cite{Yu1}
\begin{equation}
W^{-1}a_jW=(-1)^ja_j.\label{W-symmtry}
\end{equation}
Then we obtain
\begin{eqnarray}
W^{-1}H_\mathrm{H}W&&=-H_\mathrm{H},\notag\\
W^{-1}H_\mathrm{F}W&&=H_\mathrm{F},\notag\\
W^{-1}H_0W&&=H_0,
\end{eqnarray}
with the hopping $H_\mathrm{H}=-J\sum_{j}(a^\dagger_ja_{j+1}+\mathrm{H.c})$, the force $H_\mathrm{F}=F\sum_{j}ja^\dagger_ja_j$, and the nearest-neighbour interaction $H_0=V\sum_j\hat{n}_j\hat{n}_{j+1}$.
Let the Hamiltonian and initial state be invariant under the time-reversal operator $R$,  which only changes the imaginary part, $R^{-1}i R=-i$.
With the help of these operators $W$ and $R$, we have
\begin{eqnarray}
n_j(t)&&|_{(V,F)}\notag\\
&&=\langle\psi(t)|a^\dagger_ja_j|\psi(t)\rangle\notag\\
&&=\langle\psi(0)|e^{i(H_H+H_0+H_\mathrm{F})t}a^\dagger_ja_j e^{-i(H_H+H_0+H_\mathrm{F})t}|\psi(0)\rangle\notag\\
&&=\langle\psi(0)|(WR)^{-1}e^{i(H_H+H_0+H_\mathrm{F})t}RWa^\dagger_ja_j(WR)^{-1}\notag\\
&&\quad \times e^{-i(H_H+H_0+H_\mathrm{F})t}RW|\psi(0)\rangle\notag\\
&&=\langle\psi(0)|e^{i(H_H-H_0-H_\mathrm{F})t}a^\dagger_ja_je^{-i(H_H-H_0-H_\mathrm{F})t}|\psi(0)\rangle\notag\\
&&=n_j(t)|_{(-V,-F)}.
\label{EQNS1}
\end{eqnarray}
When $F=0$, it reduces to $n_j(t)|_V=n_j(t)|_{-V}$.
Similarly, defining the lattice-reversal operator
\begin{equation}
Q^{-1}a_jQ=a_{-j},
\end{equation}
we have
\begin{eqnarray}
Q^{-1}H_\mathrm{H}Q&&=H_\mathrm{H},\notag\\
Q^{-1}H_\mathrm{F}Q&&=-H_\mathrm{F},\notag\\
Q^{-1}H_0Q&&=H_0.
\end{eqnarray}
And the initial state is invariant under the operator $Q$.
Then following Eq.(\ref{EQNS1}), we obtain
\begin{eqnarray}
n_j(t)&&|_{(V,F)}\notag\\
&&=\langle\psi(0)|e^{i(H_H-H_0-H_\mathrm{F})t}a^\dagger_ja_je^{-i(H_H-H_0-H_\mathrm{F})t}|\psi(0)\rangle\notag\\
&&=\langle\psi(0)|Q^{-1}e^{i(H_H-H_0-H_\mathrm{F})t}QQ^{-1}a^\dagger_ja_jQ\notag\\
&&\quad \times Q^{-1}e^{-i(H_H-H_0-H_\mathrm{F})t}Q|\psi(0)\rangle\notag\\
&&=\langle\psi(0)|e^{i(H_H-H_0+H_\mathrm{F})t}a^\dagger_{-j}a_{-j}\notag\\
&&\quad \times e^{-i(H_H-H_0+H_\mathrm{F})t}|\psi(0)\rangle\notag\\
&&=n_{-j}(t)|_{(-V,F)}.
\end{eqnarray}
Because of these symmetries, we only need to study the dynamics for systems with attractive interactions.

\begin{figure*}[tbp]
\begin{center}
\includegraphics[scale=0.7, bb=14 291 560 572]{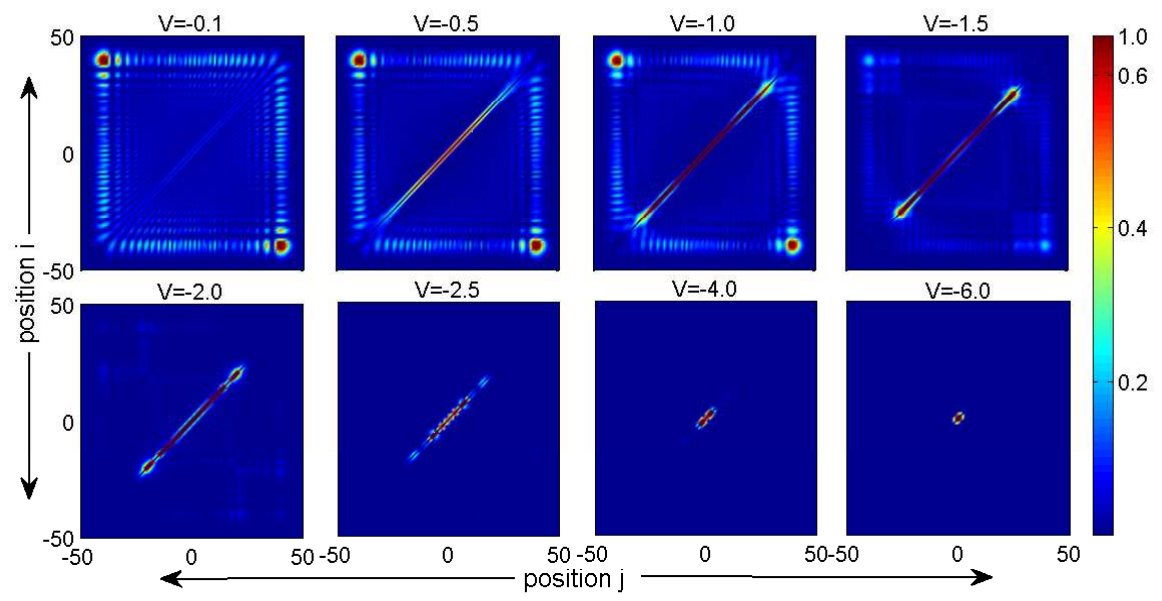}
\includegraphics[scale=0.7, bb=16 288 565 576]{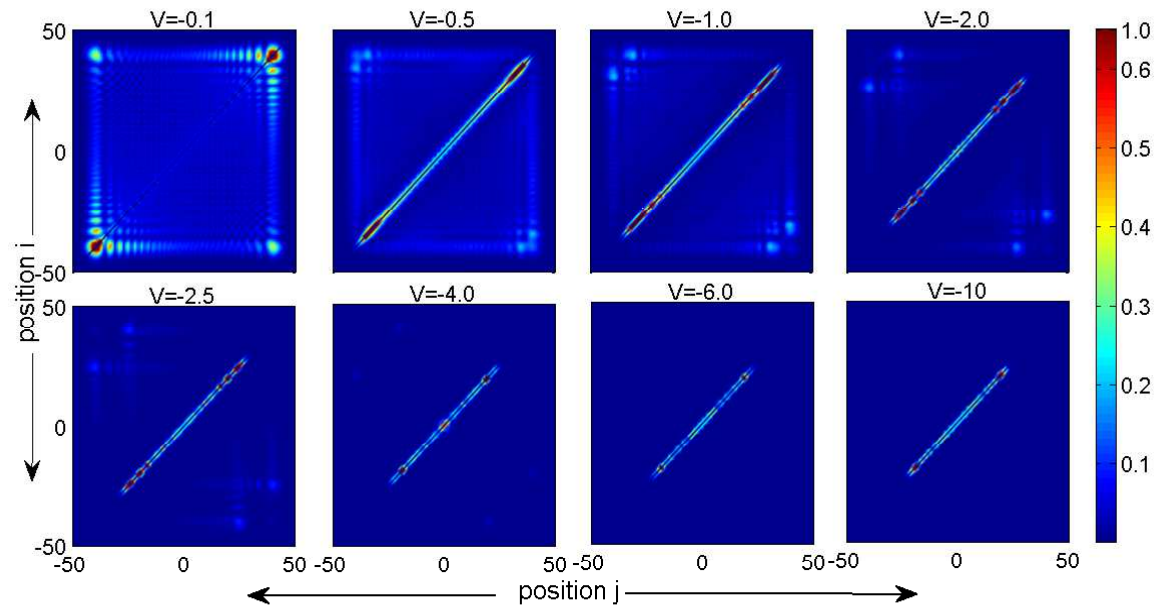}
\caption{(Color online) Density-density correlation functions $C_{i,j}=\langle\psi(t)|\hat{n}_i\hat{n}_j|\psi(t)\rangle$  for three-fermion (upper panel) and three-boson (lower panel) QWs, at the time $t=22$ when particles in scattering states are about to collide with boundaries.}
\label{FigS1}
\end{center}
\end{figure*}

\section{S2. Correlations in three-particle quantum walks}

The time-dependent density-density correlation function in real space is defined by
\begin{equation}
C_{i,j}(t)=\langle\psi(t)|\hat{n}_i\hat{n}_j|\psi(t)\rangle,
\end{equation}
which can be used to study the three-particle quantum walks (QWs).
It is very  important  to distinguish co-walking from independent walking.
In a co-walking, particles bound together and move as a single composite particle.
Significant correlations show  few specific lines ($i=j\pm d$) in the ($i,j$) plane,  which give  a signature of the co-walking of particles,
where $d$ can be integers which depend on the form of interaction and the particle number.
For three-particle systems with nearest-neighbour attractions, $d=1,2$.
In Fig.\ref{FigS1}  we show density-density correlation functions at the time $t=22$ for both fermionic and bosonic systems.
A hard-core bosonic system has the same density-density correlation function as  the one for the corresponding fermionic system.
The correlation function also has the symmetry $C_{i,j}|_V=C_{i,j}|_{-V}$, which can be proved by using  the same symmetries as being  used above.
%
%
Therefore, we demonstrate  correlations between particles in the attractive systems  in Fig.\ref{FigS1}.
At the time $t=22$, particles in scattering states (fastest) are about to reach the  boundaries so that  here the  boundary effects are negligible.

Like  the Hanbury-Brown-Twiss (HBT) interference, for  a small attraction $|V|$,  the density-density correlation function shows anti-bunching behavior with a major off-diagonal correlation at the wave front for the fermionic system, while it shows bunching behavior with  a major diagonal correlation  for the bosonic system.
A similar behaviour was found  in the two-particle QWs \cite{Qin1}.
When $|V|$ increases, bunching and anti-bunching gradually fade out, and minor diagonal correlation  lines ($i=j\pm1,2$) are enhanced in statistics-dependent QWs.
The  correlations on these diagonal lines are the  robust signature of the three-particle quantum co-walking, showing  the existence of three-particle bound states.
Whereas,  in two-particle QWs, the co-walking is characterized by the correlations on the minor diagonal lines ($i=j\pm1$) \cite{Qin1}.
For a system with a very large $|V|$ only the co-walking remains.
However, the three-particle co-walking is different from the two-particle case and essentially depends on the statistics of the particles. %
Detailed discussions about the co-walking are presented in the main text and next section.

\section{S3. Many-body perturbation and effective single-particle models}

Under a strong attraction, three-body bound states of bosons and fermions dominate significantly different   QWs and Bloch oscillations (BOs).
Three particles behave like a single composite particle and perform quantum co-walking or co-BO.
In these cases, we can treat the hopping $H_\mathrm{H}$ and additional force $H_\mathrm{F}$ as perturbations to the interaction $H_0$.
To implement the perturbation, we first give the projection operator onto the subspace involved in the three-particle co-walking (co-BO).
Then projecting the total Hamiltonian onto the subspace we can obtain effective single-particle models for QWs and BOs of three bosons and fermions.
Following the procedure shown in Ref. \cite{Takahashi1}, we first rewrite the Hamiltonian as
\begin{eqnarray}
H&=&H_0+H_1,\notag\\
H_1&=&H_\mathrm{H}+H_\mathrm{F},
\end{eqnarray}
and treat $H_1$ as a perturbation to $H_0$.
We assume that $H_0$ has a discrete degenerate level $E_0$, the subspace spanned by corresponding degenerate states is denoted by $U_0$, and the projection operator onto $U_0$ is denoted by $P_0$.
We designate the subspace of perturbed eigenstates as $U$ and its corresponding projection operator as $P$.
Here $P$ is given by the following integral of resolvent
\begin{eqnarray}
P&&=\frac{1}{2\pi i}\oint_Cdz(z-H_0-H_1)^{-1}\notag\\
&&=\frac{1}{2\pi i}\oint_Cdz(z-H_0)^{-1}\sum_{n=0}^\infty[H_1(z-H_0)^{-1}]^n.
\end{eqnarray}
Contour $C$ contains no eigenvalues of $H_0$ except $E_0$ in the complex plane.
Using $(z-H_0)^{-1}=(z-E_0)^{-1}P_0+(1-P_0)(z-H_0)^{-1}$, we obtain
\begin{equation}
P=P_0-\sum_{n=1}^\infty\sum_{k_1+k_2+...+k_{n+1}=n,k_i\geq0}S^{k_1}H_1S^{k_2}H_1...H_1S^{k_{n+1}},
\label{EQNS2}
\end{equation}
with
\begin{eqnarray}
S^0&=&-P_0,\notag\\
S^k&=&[(1-P_0)/(E_0-H_0)]^k\quad\mathrm{ for }\quad k\geq1.
\end{eqnarray}
Considering a transformation from a state $|\phi\rangle$ in $U_0$ to a state $|\psi\rangle$ in $U$:
\begin{eqnarray}
\label{EQNS3}
&&|\psi\rangle=\Gamma|\phi\rangle,\quad \Gamma\equiv PP_0(P_0PP_0)^{-1/2},\\
&&(P_0PP_0)^{-1/2}\equiv P_0+\sum_{n=1}^\infty\frac{(2n-1)!!}{(2n)!!}[P_0(P_0-P)P_0]^n.\notag
\end{eqnarray}
We can obtain
\begin{eqnarray}
&&\Gamma^\dagger\Gamma=P_0,\notag\\
&&\langle\Gamma\phi|\Gamma\phi'\rangle=\langle\phi|\phi'\rangle,
\end{eqnarray}
for any $|\phi\rangle,|\phi'\rangle\in U_0$.
Then the eigenvalue problem $H|\psi\rangle=E|\psi\rangle$ is replaced by
\begin{equation}
h|\phi\rangle=E|\phi\rangle,\quad h\equiv\Gamma^\dagger H\Gamma.
\end{equation}
$h$ is the effective Hamiltonian in $U_0$ subspace.
Next we  will use this   perturbation theory to treat QWs and BOs in the  three-particle systems.

\subsection{A. Fermions}

\begin{figure}[tbp]
\begin{center}
\includegraphics[scale=0.6, bb=136 296 480 540]{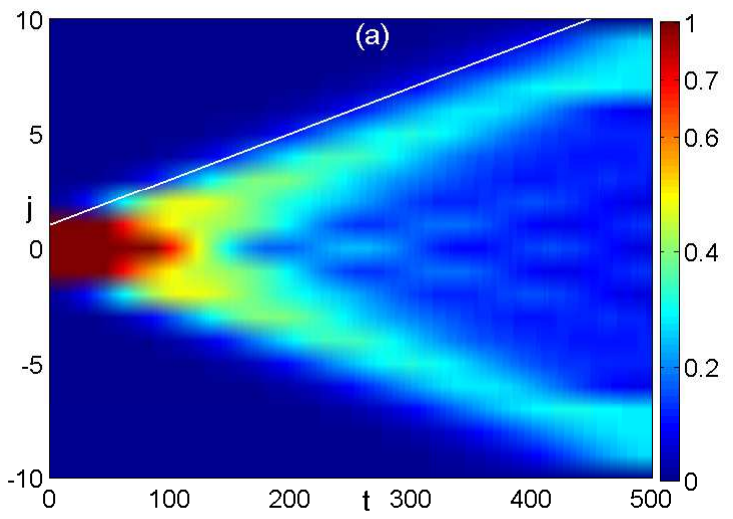}\\
\vspace{6pt}
\includegraphics[scale=0.6, bb=136 296 483 542]{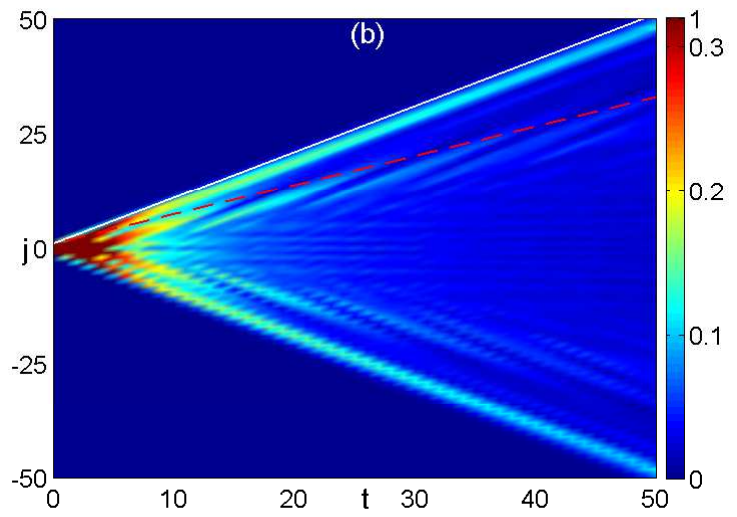}
\caption{(Color online) (a) Long-time evolution of density distribution for the fermionic system with $L=21$, $F=0$ and $V=-10$.
The straight line corresponds to the position of wave front, with a slope equaling the maximal group velocity $v^F =2/V^2$.
(b) Time evolution of density distribution for the bosonic system with $L =101$, $F=0$ and $V=-40$.
Straight lines have slopes $v^B_{1,2}\simeq1,0.64$. }
\label{FigS3}
\end{center}
\end{figure}

The unperturbed Hamiltonian $H_0$ has three eigenvalues for the three-fermion system:
\newline(i) $E_0=2V$ for the $L$-fold degenerate ground states  $\{|G_j\rangle=|j-1,j,j+1\rangle\}$, which are three-body bound states, \newline(ii) $E_{j_1j_2}=V$ for two-body bound states $\{|E_{j_1j_2}\rangle=|j_1,j_1+1,j_2\rangle: j_2\neq j_1-1,...,j_1+2\}$, \newline(iii)$E_{j_1j_2j_3}=0$ for scattering states $\{|E_{j_1j_2j_3}\rangle=|j_1,j_2,j_3\rangle: j_1+1<j_2<j_3-1\}$. \newline
$|j_i\rangle$ denotes a fermion located  at the  site $j_i$, and $L$ is the total number of lattice sites.
The three-fermion co-walking ({co-BO}) only involves the subspace spanned by $L$ independent ground states $|G_j\rangle$.
The projection operator onto the subspace is
\begin{equation}
P_0=\sum_j|G_j\rangle\langle G_j|, \label{p0}
\end{equation}
and corresponding orthogonal projection operator reads
\begin{eqnarray}
S=&&\sum_{j_1,j_2}\frac{1}{E_0-E_{j_1j_2}}|E_{j_1j_2}\rangle\langle E_{j_1j_2}|\notag\\
&&+\sum_{j_1,j_2,j_3}\frac{1}{E_0-E_{j_1j_2j_3}}|E_{j_1j_2j_3}\rangle\langle E_{j_1j_2j_3}|.\label{S0}
\end{eqnarray}
Expanding Eq.(\ref{EQNS2}) and Eq.(\ref{EQNS3}),  up to third order perturbations,  the effective Hamiltonian  reads
\begin{eqnarray}
H_{\mathrm{eff}}^{(3)}&=&E_0P_0+P_0H_1P_0+P_0H_1SH_1P_0\notag\\
&&+P_0H_1SH_1SH_1P_0\\
&&-\frac{1}{2}(P_0H_1P_0H_1S^2H_1P_0+P_0H_1S^2H_1P_0H_1P_0).\notag
\end{eqnarray}
Substituting  $P_0$ (\ref{p0}), $S$ (\ref{S0}) and $H_1$ into the above equation, we obtain the non trivial effective Hamiltonian
\begin{eqnarray}
H_{\mathrm{eff}}^{(3)}&=&h_1^{(3)}+h_2^{(3)},\notag\\
h_1^{(3)}&=&-\frac{J^3}{V^2}\sum_j(|G_j\rangle\langle G_{j+1}|+|G_{j+1}\rangle\langle G_j|),\notag\\
h_2^{(3)}&=&3F\sum_jj|G_j\rangle\langle G_{j}|.
\end{eqnarray}
$h_1^{(3)}$ and $h_2^{(3)}$ correspond to the hopping and force terms, respectively.
In order to capture the single-particle nature, we introduce creation operators $c^\dagger_j$ for states $|G_j\rangle$ ($c^\dagger_j|0\rangle=a^\dagger_{j-1}a^\dagger_ja^\dagger_{j+1}|0\rangle$).
Then the three-fermion co-walking (co-BO) obeys the following  effective single-particle Hamiltonian
\begin{equation}
\hat{H}^\mathrm{F}_{\mathrm{eff}}=-\frac{J^3}{V^2}\sum_j(c^\dagger_jc_{j+1}+\mathrm{H.c.})+3F\sum_jjc^\dagger_jc_j.
\end{equation}
For fermions, the first and second orders are trivial, because the translation of a three-fermion bound state can not be realized by hopping of one or two fermions. 
The third order is nontrivial and is the only process  for the  three-body bound state moving right.
In Fig.\ref{FigS3} (a) we show long-time evolution of density distribution for a fermionic system with $F=0$ and a very large attraction.
The  straight line has a slope $v^\mathrm{F}=2J^3/V^2$ which shows  a maximal group velocity of the effective single-particle Hamiltonian at the absence of the external force ($F=0$).
This analytical  straight line indicates  the wave front of three-fermion co-walking and is consistent with the numerical calculation.
We observe that such  consistence holds quite well  as long as $V/J\geq3$, although the perturbation theory requires only for  $J\ll V$.

\subsection{B. Bosons}

For the three-boson system, the unperturbed Hamiltonian $H_0$ also has three eigenvalues: \newline(i) $E_0=2V$ for the $3L$-fold degenerate ground states (three-body bound states) $\{|G^{(1)}_j\rangle=|j-1,j,j+1\rangle\}\cup\{|G^{(2)}_j\rangle=|j^2,j+1\rangle\}\cup\{|G^{(3)}_j\rangle=|j-1,j^2\rangle\} $, \newline(ii) $E_1=V$ for two-body bound states $\{|E_{j_1j_2}\rangle=|j_1,j_1+1,j_2\rangle: j_2\neq j_1-1,...,j_1+2\}$, \newline(iii)$E_2=0$ for scattering states $\{|D_{j_1j_2j_3}\rangle=|j_1,j_2,j_3\rangle: j_1+1<j_2<j_3-1\}\cup\{|D_{j_1j_2}\rangle=|j^2_1,j_2\rangle: j_2\neq j_1\pm1\}\cup\{|D_{j_1}\rangle=|j^3_1\rangle\}$.
Here $|j^n_1\rangle$ means that there are $n$ bosons at lattice site $j_1$.
The three-boson co-walking (co-BO) involves the subspace spanned by $3L$-fold degenerate ground states $|G^{(i=1,2,3)}_j\rangle$.
States with different $i$ or $j$ are orthogonal to each other.
The projection operator onto the subspace
\begin{equation}
P_0=\sum_j(|G^{(1)}_j\rangle\langle G^{(1)}_j|+|G^{(2)}_j\rangle\langle G^{(2)}_j|+|G^{(3)}_j\rangle\langle G^{(3)}_j|).
\end{equation}
Unlike the fermionic system, the bosonic one can have more than one particle per site.
%
%
States with different $i$ in the subspace can be transformed into each other by the perturbation Hamiltonian $H_1$.
Then the first order effective Hamiltonian is non trivial, while the zeroth order is proportional to the identity operator.
This is quite different from the three-fermion and two-particle cases \cite{Qin1}.
After a lengthy calculation the first order effective Hamiltonian reads
\begin{eqnarray}
H^{(1)}_{\mathrm{eff}}=&&P_0H_1P_0=h^{(1)}_1+h^{(1)}_2,\notag\\
h^{(1)}_1=&&-\sqrt{2}J\sum_j(|G^{(3)}_j\rangle\langle G^{(1)}_j|+|G^{(2)}_j\rangle\langle G^{(1)}_j|\notag\\
&&+\sqrt{2}|G^{(3)}_{j+1}\rangle\langle G^{(2)}_j|+\mathrm{H.c.})\notag\\
h^{(1)}_2=&&F\sum_j3j|G^{(1)}_j\rangle\langle G^{(1)}_j|+F\sum_j(3j+1)|G^{(2)}_j\rangle\langle G^{(2)}_j|\notag\\
&&+F\sum_j(3j-1)|G^{(3)}_j\rangle\langle G^{(3)}_j|.
\end{eqnarray}
Then we introduce effective single-particle creation operators.
Explicitly, $b^\dagger_j|0\rangle=a^\dagger_{j-1}a^\dagger_{j}a^\dagger_{j+1}|0\rangle$, $c^\dagger_j|0\rangle=\frac{1}{\sqrt{2}}(a^\dagger_{j})^2a^\dagger_{j+1}|0\rangle$, and $d^\dagger_j|0\rangle=\frac{1}{\sqrt{2}}a^\dagger_{j-1}(a^\dagger_{j})^2|0\rangle$.
The effective single-particle Hamiltonian for three-boson co-walking (co-BO) is given by
\begin{eqnarray}
H^\mathrm{B}_\mathrm{eff}=&&-\sqrt{2}J\sum_j(d^\dagger_jb_j+c^\dagger_jb_j+\sqrt{2}d^\dagger_{j+1}c_j+\mathrm{H.c.})\notag\\
&&+F\sum_j3jb^\dagger_jb_j+F\sum_j(3j+1)c^\dagger_jc_j\notag\\
&&+F\sum_j(3j-1)d^\dagger_jd_j.
\end{eqnarray}
It is independent of $V$.
The amplitudes are proportional to $J$, while amplitudes of nontrivial higher orders are much smaller than the leading order term  when $J/V\ll1$. 
 On the other hand, the first order perturbation of the force term is also nontrivial.
  The force can not transfer states in the three-body bound state subspace. 
  It thus  is diagonal and only causes energy shifts of different states.

After transforming it into the momentum space, the hopping part of the effective single-particle Hamiltonian reads
\begin{equation}
H^\mathrm{B}_\mathrm{eff}=-J\sum_k\Psi^\dagger_k
\left(
\begin{matrix}
0&\sqrt{2}&\sqrt{2}&\\
\sqrt{2}&0&2e^{-ik}&\\
\sqrt{2}&2e^{ik}&0&
\end{matrix}
\right)
\Psi_k.
\label{EQNS4}
\end{equation}
with $(\Psi_k)^\mathrm{T}=(b_k,c_k,d_k)$.
We see that analytical eigenvalues are quite complicated.
Therefore we just presented numerical results in the main text.

The Hamiltonian Eq.(\ref{EQNS4}) consists with the  three mini-band spectra.
The maximal group velocity for the middle mini-band is about $v^\mathrm{B}_1\simeq1$, and the other two mini-bands have the same maximal group velocity $v^\mathrm{B}_2\simeq0.64$.
In Fig.\ref{FigS3} (b) we show the time evolution of density distribution for a three-boson system with $F=0$ and a very large attraction.
The white straight and red dot lines show  slopes of  $v^\mathrm{B}_1$ and $v^\mathrm{B}_2$, respectively.
They truly indicate  the  movements of the wave fronts for two inner light cones when $V/J\geq6$, revealing an important  signature of  the three-boson co-walking.

\section{S4. Two-body bound states  in three-particle quantum walks}

In this section, we demonstrate that two-body physics can emerge in three-particle QWs and BOs.
The quantum walks of two particles is quite contrast to the situation in the quantum walks of three particles.
 The co-BO states  are also quite different in the two-body bound state sector.
To this end, we employ the initial state $|\psi_1(t=0)\rangle=a^\dagger_{-1}a^\dagger_0a^\dagger_2|0\rangle$, with one particle being away from the other neighbouring two.
The initial state is an eigenstate in 2BS subspace of the interaction term $H_0$.
Similar symmetries in dynamics still hold, and we only study systems with attractive interactions.

\subsection{A. Quantum walks}

\begin{figure}[tbp]
\begin{center}
\includegraphics[width=\linewidth, bb=152 280 443 616]{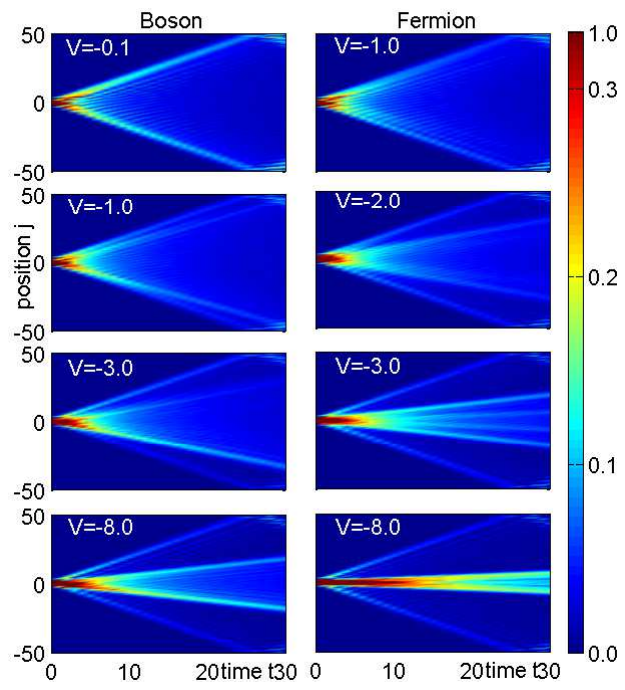}
\caption{(Color online) Time evolutions of density distribution for three-boson (left collum) and fermion (right collum) QWs, from the initial state $|\psi_1\rangle=a^\dagger_{-1}a^\dagger_0a^\dagger_2|0\rangle$. The inner-cones show the QWs of the two-body bound states. }
\label{Fig2BSQW}
\end{center}
\end{figure}

\begin{figure}[tbp]
\begin{center}
\includegraphics[width=\linewidth, bb=91 306 500 540]{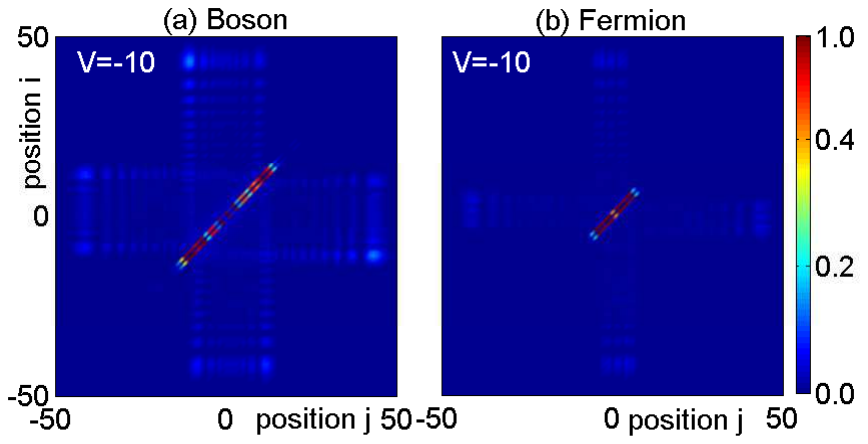}
\caption{(Color online) Density-density correlation functions $C_{i,j}$ at the time $t=22$, for three-boson (a) and fermion (b) QWs from the initial state $|\psi_1\rangle=a^\dagger_{-1}a^\dagger_0a^\dagger_2|0\rangle$. For a strong  interaction, i.e., $V=-10$,  the correlation functions mainly contain two diagonal lines at $i=j\pm1$ in the ($i,j$) plane, showing the presence of two-particle bound states. The big ``Cross" in the correlation for bosons clearly  shows the interference between the pair and single particle.  }
\label{Fig2BSCorr}
\end{center}
\end{figure}

In Fig.\ref{Fig2BSQW} we show time evolutions of density distribution for three-particle QWs from the initial state $|\psi_1\rangle$.
In the non-interacting limit, both bosonic and fermionic systems show single-particle nature and experience a ballistic expansion with speed of wave front (SWF) $2$.
For a bosonic system with non-zero attraction, there is an inner light-cone and corresponding SWF decreases as the attraction increases.
While for the fermionic system, the inner light-cone exists only when $|V|>1$, and corresponding SWF has a different decay pace from the bosonic one.
Compared with QWs from the initial state $|\psi\rangle=a^\dagger_{-1}a^\dagger_0a^\dagger_1|0\rangle$ shown in the main paper (see also in Fig.\ref{FigS6} and Fig.\ref{FigS7} in last section), here the outer light-cone does not fade out as $|V|$ increases, and a third innermost light-cone shows up for systems with intermediate attractions but disappears at large attractions.
The expansion contains two light-cones for systems with very large attractions, and the SWF of inner light-cone for bosons is larger than that for fermions.

The density-density correlation function for QWs from the initial state $|\psi_1\rangle$ shows similar behaviours as for the initial state $|\psi\rangle$, when the attraction is not very large.
(Anti)bunching behavior for bosonic (fermionic) systems with weak attractions fades out as the attraction increases, along with the gradually enhanced diagonal correlations.
In Fig.\ref{Fig2BSCorr} we show correlation functions for systems with a large attraction.
Both bosonic and fermionic systems have similar behaviours.
Correlation functions mainly contain two diagonal lines at $i=j\pm1$ in the ($i,j$) plane, which indicates the presence of two-particle bound states.
It also presents a characteristic of two-particle quantum co-walking, see  \cite{Qin1}.
%
We observe  the two-particle pair in inner light-cone and the 'free' particle in outer light-cone in time evolutions. %
While the interference between the single particle and the pair is visible in Fig.\ref{Fig2BSCorr}.
The big ``Cross" in the correlation for bosons clearly  shows the interference between the pair and single particle.
But such an interference  is much weaker in the QWs of  the three  fermions.
In the following, we will derive an effective Hamiltonian from which we can conceive such subtle dynamics of two-body bound states in the QWs of three particles.
Under a strong attraction, an effective single-particle model also can be derived for QWs from the initial state $|\psi_1\rangle$.
After projecting onto the subspace of 2BSs, the second order perturbation gives
\begin{equation}
H^{F(B)}_{\mathrm{eff}}=H_p^{F(B)}+H_s+H_t
\label{EFF}
\end{equation}
with
\begin{eqnarray}
H_p^{F(B)}=&&\frac{(3)J^2}{V}\sum_{j_1,j_2}(a^\dagger_{j_1-1,j_2}a_{j_1j_2}+\mathrm{h.c.}),\\
H_s=&&-J\sum_{j_1,j_2}(a^\dagger_{j_1,j_2+1}a_{j_1j_2}+\mathrm{h.c.}),\\
H_t=-&&J\sum_j(a^\dagger_{j-2,j+1}a_{j,j-2}+a^\dagger_{j+2,j}a_{j,j+3})\notag\\
-&&\frac{J^2}{V}\sum_j(a^\dagger_{j+2,j-1}a_{j,j+3}+a^\dagger_{j-2,j+2}a_{j,j-2})\notag\\
+&&\frac{J^2}{V}\sum_j(a^\dagger_{j-3,j+1}a_{j,j-3}+a^\dagger_{j+1,j-1}a_{j,j+3})\notag\\
+&&\frac{J^2}{V}\sum_j(a^\dagger_{j-2,j+1}a_{j,j-3}+a^\dagger_{j+2,j-1}a_{j,j+3})\notag\\
+&&\frac{J^2}{V}\sum_j(a^\dagger_{j+2,j}a_{j,j+4}+a^\dagger_{j-2,j+2}a_{j,j-2})\notag\\
+&&\frac{J^2}{V}\sum_j(a^\dagger_{j+3,j}a_{j,j+4}+a^\dagger_{j-1,j+2}a_{j,j-2}).
\end{eqnarray}
In the above equations, we defined effective single-particle operators $a^\dagger_{j_1,j_2}|0\rangle=|E_{j_1,j_2}\rangle$ for 2BSs, where $j_1$ and $j_2$ denote positions of the two-particle pair and the 'free' particle away from the pair, respectively.
The summation on $j_2$ is restricted in the set $\{j_2\neq j_1-2,..,j_1+2\}$.
 ``F(B)" stands for the fermionic (bosonic) system.
The first term $H_p^{F(B)}$ indicates the hoping occurs in the paired particle and the second term $H_s$  shows the hopping occurs in the  single particle.
While the last term $H_t$ shows the interchange process between the pair and single particle.
Moreover, the effective Hamiltonians $H_p^{F(B)}$ govern the dynamics of two-particle pair, which are the same as the Hamiltonians for two-particle quantum co-walkings \cite{Qin1}, where the spectra of $H_p^{F(B)}$ read $E_p^{F(B)}(K)=\frac{(3)2J^2}{V}\mathrm{cos}(K)$.
The maximal group velocities $\upsilon_p^{F(B)}=(3)2J^2/V$, which determine the SWFs of the inner light-cone.
The bosonic pair expanses three times faster than the fermionic one.
Unlike for the three-particle quantum co-walkings, maximal group velocities for both systems decrease quickly as $|V|$ increases.
Hamiltonian $H_s$ governs the dynamics of the 'free' particle in 2BSs, which is the same as the single-particle Hamiltonian with a maximal group velocity $2J$.
Finally, the  effective Hamiltonian $H_t$ describes interference between the two-particle pair and the 'free' particle, see the Fig.\ref{Fig2BSCorr}.

\subsection{B. Bloch oscillations}

\begin{figure}[tbp]
\begin{center}
\includegraphics[width=\linewidth, bb=153 266 443 653]{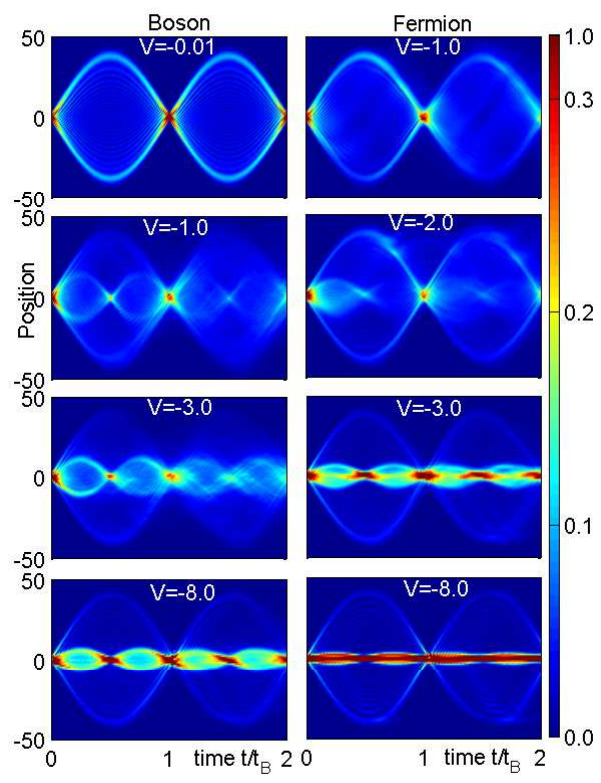}
\caption{(Color online) Time evolutions of density distribution for three-boson (left column) and fermion (right column) BOs from the initial state $|\psi_1\rangle$. $L=101$ and $F=0.1$. Inner cycles show the two-body bound state BOs in the QWs of three particles. }
\label{Fig2BSBO}
\end{center}
\end{figure}

%
In this subsection, we study BOs from the initial state $|\psi_1\rangle$.
The Hamiltonian of  an  additional force is the same ($H_\mathrm{F}=F\sum_jja^\dagger_ja_j$) for both bosonic and fermionic systems.
In Fig.\ref{Fig2BSBO} we show BOs for both bosonic and fermionic systems with different attractions.
For fermionic systems with $|V|<1$ and non-interacting bosons, the particles show  the single-particle BO independently, with an oscillation amplitude $4J/F$ and a period $t_B=2\pi/F$.
As the attraction increases, the single-particle BO does not fade out.
A gradually enhanced inner BO appears with a temporal period $t_B/2$.
Its amplitude decreases as the attraction increases and has different decay paces for bosons and fermions.
For  a large attraction, the amplitude of inner BO for bosons is much  larger than that for fermions.
Under large attractions, effective single-particle models for BOs from the initial state $|\psi_1\rangle$ are
\begin{equation}
H_\mathrm{BO}^{F(B)}=H^{F(B)}_{\mathrm{eff}}+H_{1F}+H_{2F},
\end{equation}
with
\begin{eqnarray}
H_{1F}&=&F\sum_{j_1,j_2}j_2a^\dagger_{j_1,j_2}a_{j_1,j_2},\notag\\ H_{2F}&=&F\sum_{j_1,j_2}(2j_1+1)a^\dagger_{j_1,j_2}a_{j_1,j_2}.
\end{eqnarray}
They are derived by up to second order perturbations.
$H^{F(B)}_{\mathrm{eff}}$ were presented in Eq.(\ref{EFF}).
$H_s+H_F$ governs the BO of 'free' particle in 2BSs, which are  the same as the Hamiltonian for single-particle BO.
$H_p^{F(B)}+H_{2F}$ describes the effective single-particle behavior of the two-particle pair in 2BSs.
The two-particle pair undergoes a co-BO with the amplitude $(3)4J^2/(VF)$ and the period $t_B/2$.
Except the three times relation in amplitudes of the co-BOs, bosons and fermions have the same behavior and no Landau-Zener tunnelling exists.

\section{S5. Quantum Fisher information}

According to Ref.\cite{Liu1}, quantum Fisher information for an unitary process $U=e^{-itH}$ from a pure initial state $|\psi_0\rangle$ is given by
\begin{equation}
\mathcal{F}=4\left\{\left[\frac{\partial}{\partial F}\langle\psi(t)|\right]\frac{\partial}{\partial F}|\psi(t)\rangle-|\langle\psi(t)|\frac{\partial}{\partial F}|\psi(t)\rangle|^2\right\},
\label{FI}
\end{equation}
where $|\psi(t)\rangle=e^{-itH}|\psi_0\rangle$. The force $F$ in Hamiltonian $H$ is to be the measured parameter in the precision measurement of the gravitational force.
The limit on the precision of measuring the parameter $F$ is described by the quantum Cram\'er-Rao bound \cite{Liu1}
\begin{eqnarray}
\Delta F\geq\frac{1}{\sqrt{\mathcal F}}.
\end{eqnarray}
The larger quantum Fisher information indicates a higher measurement precision.
Therefore, we can use quantum Fisher information to quantify the performance of different processes in the precision measurement of gravity.

We consider the effective single-particle Hamiltonian
\begin{equation}
H=J'\sum_j[c^\dagger_jc_{j+1}+\mathrm{h.c.}]+F'\sum_jc^\dagger_jc_j,
\end{equation}
with
\begin{eqnarray}
J'&=&-J,F'=F, \mathrm{for \thinspace single-particle \thinspace BO};\\
J'&=&3J^2/V,F'=2F, \mathrm{for \thinspace two-boson \thinspace co-BO};\notag\\
J'&=&J^2/V,F'=2F, \mathrm{for \thinspace two-fermion \thinspace co-BO};\notag\\
J'&=&-J^3/V^2,F'=3F, \mathrm{for \thinspace three-fermion \thinspace co-BO},\notag
\end{eqnarray}
and the one for three-boson co-BO
\begin{eqnarray}
H=&&-\sqrt{2}J\sum_j[d^\dagger_jc_j+b^\dagger_jc_j+\sqrt{2}d^\dagger_{j+1}b_j+h.c.]\\
&&+3F\sum_j[jc^\dagger_jc_j+(j+\frac{1}{3})b^\dagger_jb_j+(j-\frac{1}{3})d^\dagger_jd_j].\notag
\end{eqnarray}
Pure initial states for (co-)BOs are in the same form $|\psi_0\rangle=c^\dagger_0|0\rangle$.

In order to calculate Eq.(\ref{FI}), we first introduce some useful formulaes. %
Given $\rho=e^G$ which is a function of $F$, we have
\begin{eqnarray}
&&\frac{\partial\rho}{\partial F}=\int^1_0e^{sG}(\partial_FG)e^{(1-s)G}ds,\notag\\
&&e^GAe^{-G}=e^{\mathrm{ad}_G}(A),
\end{eqnarray}
where $\mathrm{ad}_G$ is the adjoint superoperator, i.e., $\mathrm{ad}_G(A)=[G,A]$.
Then we have
\begin{eqnarray}
(\partial_F\rho)\rho^{-1}&=&h[\mathrm{ad}_G](\partial_FG),\\
\rho^{-1}(\partial_F\rho)&=&g[\mathrm{ad}_G](\partial_FG)
\end{eqnarray}
with operator  functions $h[x]=(e^x-1)/x$ and $g[x]=(1-e^{-x})/x$.
Given above introduced formulaes, the expression of quantum Fisher information reduces to
\begin{equation}
\mathcal{F}=4t^2\{\langle\psi_0|[H^Q(t)]^2|\psi_0\rangle-\langle\psi_0|H^Q(t)|\psi_0\rangle^2\},
\end{equation}
where we have introduced a time-dependent effective Hamiltonian
\begin{equation}
H^Q(t)=h[it\cdot\mathrm{ad}_H](\partial_FH).
\end{equation}
In numerical calculations, we treat the Hamiltonian as a vector in the space spanned by the basis $\mathfrak{C}=\{|j,j'\rangle\equiv c^\dagger_jc_{j'},j,j'=-M,...,M\}$.
Then the adjoint superoperator acts as an operator in the space $\mathfrak{C}$.
Diagonalizing the adjoint operator, we can numerically get the effective Hamiltonian and quantum Fisher informations which are shown in Fig. 5 in the main paper.
$\mathcal{F}/4t^2$ (or the fluctuation of time-dependent effective Hamiltonian $\Delta H^Q$)  increases monotonously and is approximately linear dependence of the time $t$  for the time $t<t_0$.
Whereas  it approaches to a constant with damping oscillations for  $t>t_0$. 
Regarding to  the lowest order approximation, it can be approximated by the piece-wise function, which shows a linear  dependence of the time $t$ when $t<t_0$ and  approaches to a constant when $t>t_0$.
Thus we have the Fisher information $\mathcal{F}=\alpha t^3$ when $t<t_0$, while $\mathcal{F}=At^2$ when $t>t_0$. 
The coefficients $\alpha$ are extracted by numerical fitting.

The  BOs of ideal Bose gas  can be harnessed for the precision measurement of gravity \cite{Ferrari1,Tarallo1}.
 For our model in the weakly interacting region, SSs dominate the BO with the frequency $\omega_B=F/h$ and particles experience force $F$ which is proportional to the gravitational force.
For the strongly interacting region 3BSs dominate the BO with the frequency $\omega_B=3F/h$  and the three-particle formed composite quasiparticle experiences an effective force $3F$.
Such  exact three times relation between themeasurement of gravity frequencies of the BOs of single and three bosons enables  a high precision testing  EP.
Here the precision is  free from the mass ratio uncertainty of the two  isotopes in addition to  the precision limit $1/t^{3/2}$, beyond the fundamental limit  $1/t$.

\section{S6. Additional figures}

\begin{figure*}[bp]
\begin{center}
\includegraphics[width=\linewidth, bb=0 0 310 141]{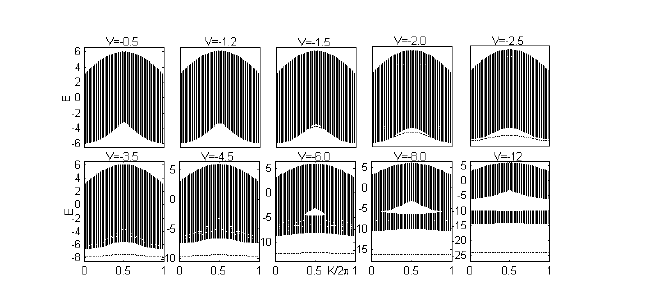}%
\caption{Three-fermion spectra for systems  with $L=61$, $F=0$ and different interaction strengths of  $V$.}
\label{FigS4}
\end{center}
\end{figure*}

\begin{figure*}[tbp]
\begin{center}
\includegraphics[width=\linewidth, bb=0 0 310 141]{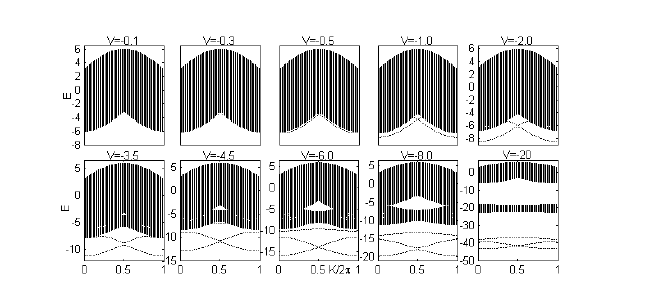}%
\caption{Three-boson  spectra for systems with $L=61$, $F=0$ and different interaction strengths.}
\label{FigS5}
\end{center}
\end{figure*}

In this subsection, we show additional figures in order to further understand the  spectra, QWs and BOs of three interacting particles.
In  Fig.\ref{FigS4} and Fig.\ref{FigS5},  we show spectra of three fermions (Fig. \ref{FigS4}) and bosons  (Fig.\ref{FigS5}) at different interaction strengths.
The figures  are obtained from  the numerical exact diagonalization in three-particle Hilbert space.
Periodic boundary conditions and the conservation of total momentum $K$ are used in the calculations.
For three bosons, the Hilbert space is spanned by the basis
\begin{equation}
\mathfrak{B}=\{|k_1k_2k_3\rangle=N_{k_1k_2k_3}a^\dagger_{k_1}a^\dagger_{k_2}a^\dagger_{k_3}|0\rangle\},\notag
\end{equation}
with $0\leq k_1\leq k_2\leq k_3< L$.
Here $N_{k_1k_2k_3}$ is the normalization factor
and $a^\dagger_k$ is the creation operator of a particle with momentum $k\frac{2\pi}{L}$.
For three fermions, the Hilbert space is spanned by the basis
\begin{equation}
\mathfrak{F}=\{|k_1k_2k_3\rangle=a^\dagger_{k_1}a^\dagger_{k_2}a^\dagger_{k_3}|0\rangle\},\notag
\end{equation}
with $0\leq k_1< k_2< k_3< L$.
Each base vector has a total momentum $K=\mathrm{mod}(\sum_{l=1}^3k_l,L)*\frac{2\pi}{L}$.
We classify the three-particle Hilbert space into a series of subspaces, each with a given total momentum  $K$.
After transforming the Hamiltonian into the momentum space, it is very easy to numerically generate the matrix representation in each subspace.
Then we can diagonalize the Hamiltonian for a pretty large size $L$.
Fig.\ref{FigS4} and Fig.\ref{FigS5} show important statistics-and-interaction-dependent nature in  many-particle spectra.

\begin{figure*}[tbp]
\begin{center}
\includegraphics[width=\linewidth, bb=14 298 566 549]{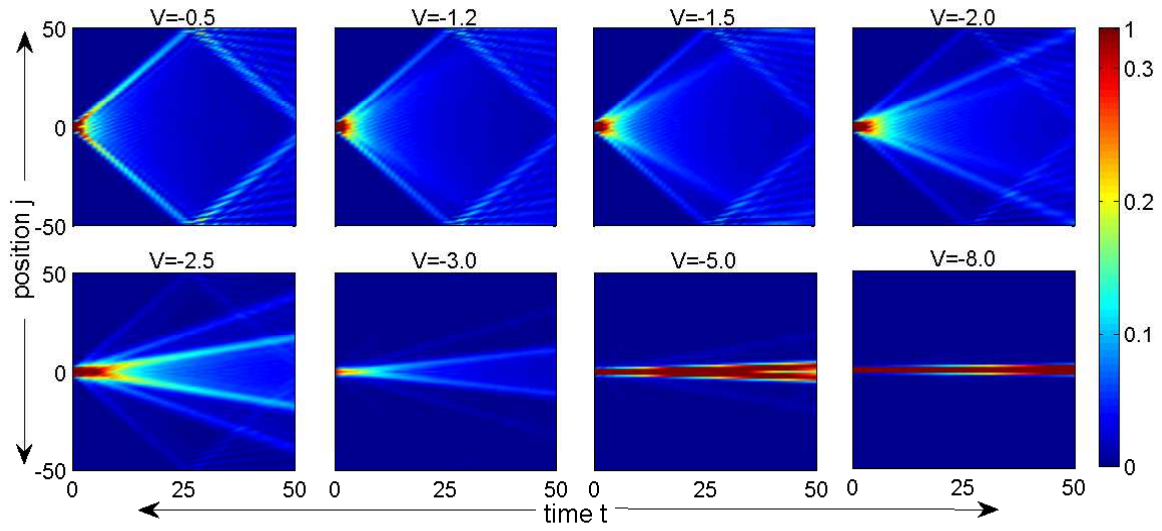}
\caption{(Color online) Time evolutions of density distribution for three-fermion systems with $L =101$, $F=0$ and different attraction strengths. }
\label{FigS6}
\end{center}
\end{figure*}

\begin{figure*}[tbp]
\begin{center}
\includegraphics[width=\linewidth, bb=15 304 560 548]{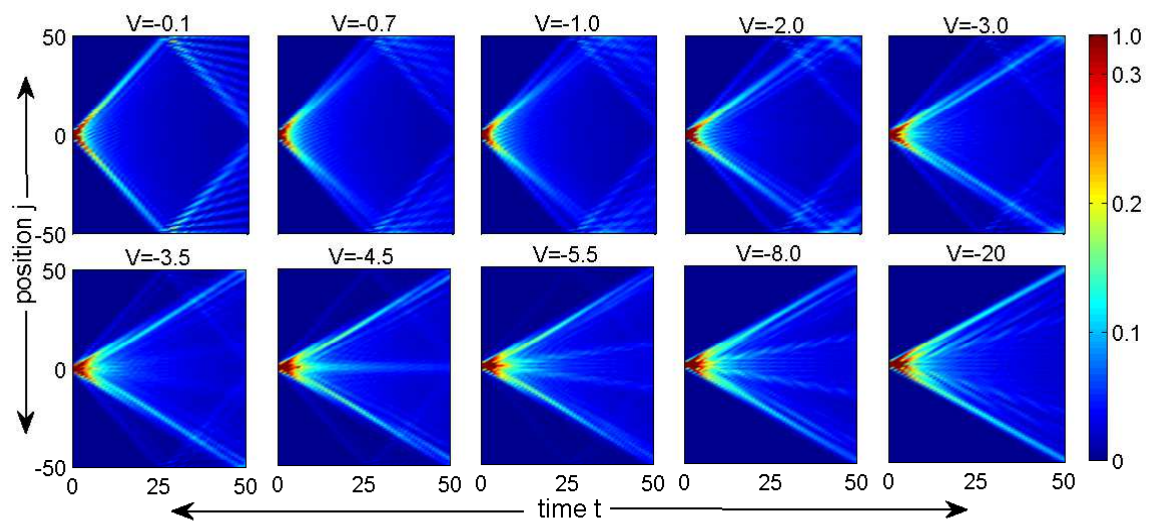}
\caption{(Color online) Time evolutions of density distribution for three-boson systems with $L =101$, $F=0$ and different attraction strengths.}
\label{FigS7}
\end{center}
\end{figure*}

In Fig.\ref{FigS6} and Fig.\ref{FigS7}, we show time evolutions of density distribution for three-fermion (Fig.\ref{FigS6}) and three-boson (Fig.\ref{FigS7}) systems with $L=101$, $F=0$  at different interaction strengths, where  the initial state is set up  $|\psi(t=0)\rangle=a^\dagger_{-1}a^\dagger_{0}a^\dagger_{1}|0\rangle$.
Time evolving block decimation (TEBD) algorithm is used.
The maximal bond dimension of matrix product state $\chi=120$.
Fifth-order Suzuki-Trotter expansion under open boundary condition is used for the unitary time evolution, and the time step $\delta t=10^{-3}$.
Total cumulative truncation errors are controlled on or less than the order of $10^{-8}$.
For systems with same parameters but different $L$, the numerical results show finite-size and boundary effects.
Such effects  are negligible before particles collide with boundaries.

\begin{figure*}[tbp]
\begin{center}
\includegraphics[scale=1.2, bb=129 201 455 716]{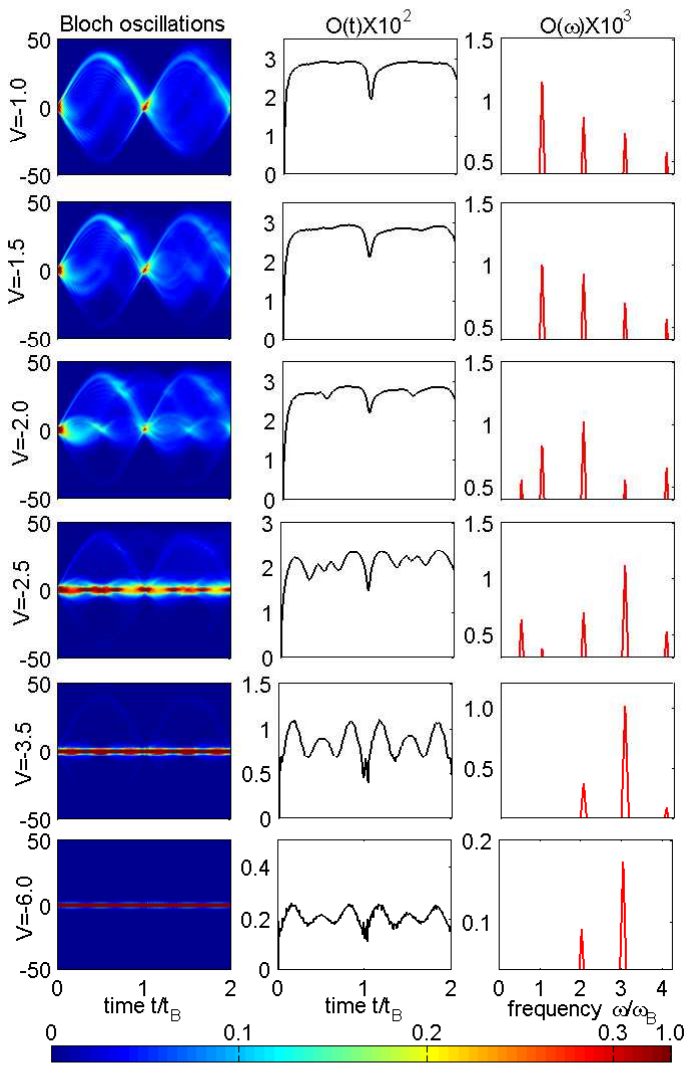}
\caption{(Color online) Time evolutions of density distribution (left column), and corresponding $O(t)$ (middle column) and $O(\omega)$ (right column) for three-fermion systems with $L =101$, $F=0.1$ and different attraction strengths.}
\label{FigS8}
\end{center}
\end{figure*}

\begin{figure*}[tbp]
\begin{center}
\includegraphics[scale=1.2, bb=128 192 448 715]{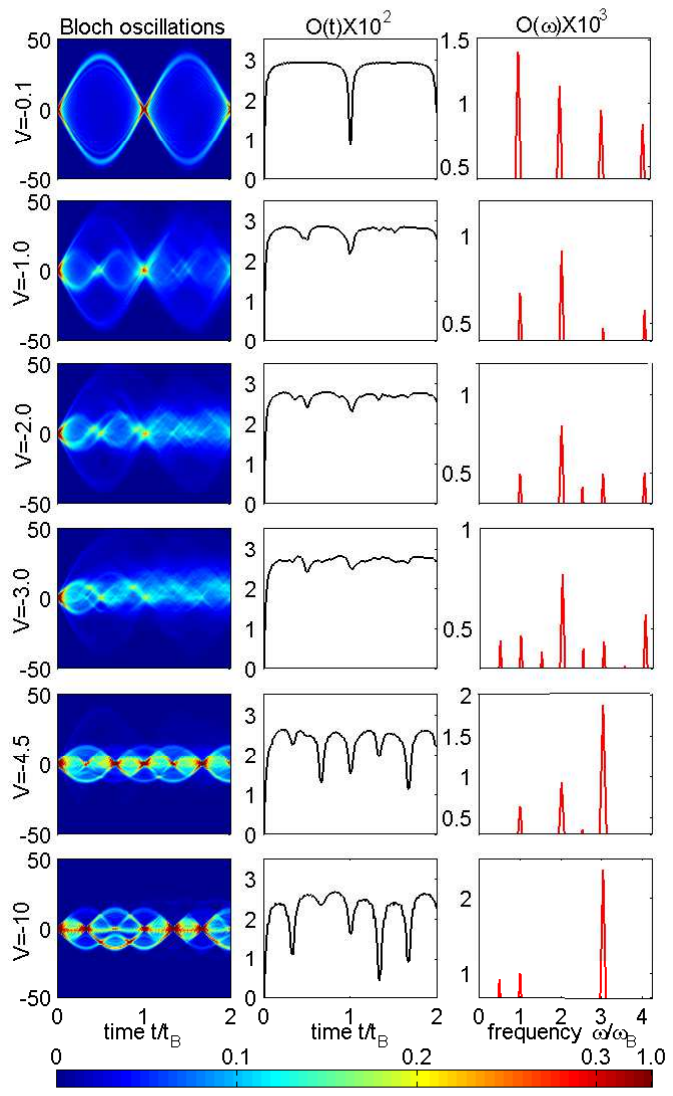}
\caption{(Color online) Time evolutions of density distribution (left column), and corresponding $O(t)$ (middle column) and $O(\omega)$ (right column) for three-boson systems with $L =101$, $F=0.1$ and different attraction strengths.}
\label{FigS9}
\end{center}
\end{figure*}

In Fig.\ref{FigS8} and Fig.\ref{FigS9},  we show time evolutions of density distribution and corresponding $O(t)$ and $O(\omega)$ for three-fermion (Fig.\ref{FigS8}) and three-boson (Fig.\ref{FigS9}) systems with $L=101$, $F=0.1$ and different attractions.
The maximal bond dimension of matrix product state $\chi=150$.
The time step for unitary time evolution is set up with $\delta t=10^{-3}t_B$, where $t_B=2\pi/F$ is the period of single-particle BO.
Total cumulative truncation errors are controlled on or less than the order of $10^{-7}$.
The statistics-dependent BOs  are demonstrated in the dynamics of three-particle systems.

\begin{references}

\bibitem{Aharonov1} Y. Aharonov, L. Davidovich, and N. Zagury, Phys. Rev. A \textbf{48}, 1687 (1993).

\bibitem{Kempe1} J. Kempe, Contemporary Phys. \textbf{44}, 307 (2003).

\bibitem{Ambainis1} A. Ambainis, Int. J. Quantum. Inform. \textbf{1}, 507 (2003).

\bibitem{Childs1} A. M. Childs, D. Gosset, and Z. Webb, Science \textbf{339}, 791 (2013).

\bibitem{Zatelli:2020} F. Zatelli, C. Benedetti, and M. G. A. Paris, arXiv:2010.12448.

\bibitem{Andraca1} S. E. Venegas-Andraca, Quant. Info. Proc. \textbf{11}, 1015 (2012).

\bibitem{Asboth1} J. K. Asb\'{o}th, Phys. Rev. B \textbf{86}, 195414 (2012).

\bibitem{Lloyd1} S. Lloyd, J. Phys. Conf. Ser. \textbf{302}, 012037 (2011).

\bibitem{Wang1} J. Wang and K. Manouchehri, \textit{Physical Implementation of Quantum Walks} (Springer, Berlin, Heidelberg, 2013).

\bibitem{Kitagawa1} T. Kitagawa, M. A. Broome, A. Fedrizzi, M. S. Rudner, E. Berg, I. Kassal, A. Aspuru-Guzik, E. Demler, and A. G. White, Nat. Commun. \textbf{3}, 882 (2012).

\bibitem{Kraus1} Y. E. Kraus, Y. Lahini, Z. Ringel, M. Verbin, and O. Zilberberg, Phys. Rev. Lett. \textbf{109}, 106402 (2012); M. Verbin, O. Zilberberg, Y. E. Kraus, Y. Lahini, and Y. Silberberg, Phys. Rev. Lett. \textbf{110}, 076403 (2013).

\bibitem{Ramasesh1} V. V. Ramasesh, E. Flurin, M. Rudner, I. Siddiqi, and N. Y. Yao, Phys. Rev. Lett. \textbf{118}, 130501 (2017).


\bibitem{Destri1} C. Destri, and H. J. De Vega, Nucl. Phys. B \textbf{290}, 363 (1987).

\bibitem{Cedzich1} C. Cedzich, T. Ryb\'{a}r, A. H. Werner, A. Alberti, M. Genske, and R. F. Werner, Phys. Rev. Lett. \textbf{111}, 160601 (2013).

\bibitem{Bisio1} A. Bisio, G. M. D'Ariano, P. Perinotti, and A. Tosini, Phys. Rev. A \textbf{97}, 032132 (2018).

\bibitem{Arnault1} P. Arnault, B. Pepper, and A. P\'{e}rez, Phys. Rev. A \textbf{101}, 062324 (2020).



\bibitem{Ahlbrecht1} A. Ahlbrecht, A. Alberti, D. Meschede, V. B. Scholz, A. H. Werner, and R. F. Werner, New J. Phys. \textbf{14}, 073050 (2012).

\bibitem{Fukuhara1} T. Fukuhara, P. Schau{\ss}, M. Endres, S. Hild, M. Cheneau, I. Bloch, and C. Gross,  (London) \textbf{502}, 76 (2013).

\bibitem{Karski1} M. Karski, L. F\"{o}rster, J.-M. Choi, A. Steffen, W. Alt, D. Meschede, and A. Widera, Science \textbf{325}, 174 (2009).

\bibitem{Schmitz1} H. Schmitz, R. Matjeschk, C. Schneider, J. Glueckert, M. Enderlein, T. Huber, and T. Schaetz, Phys. Rev. Lett. \textbf{103}, 090504 (2009); F. Z\"{a}hringer, G. Kirchmair, R. Gerritsma, E. Solano, R. Blatt, and C. F. Roos, Phys. Rev. Lett. \textbf{104}, 100503 (2010).

\bibitem{Schreiber1} A. Schreiber, K. N. Cassemiro, V. Poto\v{c}ek, A. G\'{a}bris, P. J. Mosley, E. Andersson, I. Jex, and C. Silberhorn, Phys. Rev. Lett. \textbf{104}, 050502 (2010); M. A. Broome, A. Fedrizzi, B. P. Lanyon, I. Kassal, A. Aspuru-Guzik, and A. G. White, Phys. Rev. Lett. \textbf{104}, 153602 (2010); A. Schreiber, K. N. Cassemiro, V. Poto\v{c}ek, A. G\'{a}bris, I. Jex, and C. Silberhorn, Phys. Rev. Lett. \textbf{106}, 180403 (2011).

\bibitem{Fukuhara2} T. Fukuhara, A. Kantian, M. Endres, M. Cheneau, P. Schau{\ss}, S. Hild, D. Bellem, U. Schollw\"{o}ck, T. Giamarchi, C. Gross, I. Bloch, and S. Kuhr, Nat. Phys. \textbf{9}, 235 (2013).

\bibitem{Du1} J. Du, H. Li, X. Xu, M. Shi, J. Wu, X. Zhou, and R. Han, Phys. Rev. A \textbf{67}, 042316 (2003).

\bibitem{Peruzzo1} A. Peruzzo , M. Lobino, J. C. F. Matthews, N. Matsuda, A. Politi, K. Poulios, X. Zhou, Y. Lahini, Nur Ismail, K. W\"{o}rhoff, Y. Bromberg, Y. Silberberg, M. G. Thompson, J. L. OBrien, Science \textbf{329}, 1500 (2010).
\bibitem{Mayer1} K. Mayer, M. C. Tichy, F. Mintert, T. Konrad, and A. Buchleitner Phys. Rev. A \textbf{83}, 062307 (2011).x
\bibitem{Brown1} R. H. Brown and R. Q. Twiss, Nature (London) \textbf{177}, 27 (1956).



\bibitem{Hillery1} M. Hillery, Science \textbf{329}, 1477 (2010).

\bibitem{Sansoni1} L. Sansoni, F. Sciarrino, G. Vallone, P. Mataloni, A. Crespi, R. Ramponi, and R. Osellame, Phys. Rev. Lett. \textbf{108}, 010502 (2012).

\bibitem{Solntsev1} A. S. Solntsev, A. A. Sukhorukov, D. N. Neshev, and Y. S. Kivshar, Phys. Rev. Lett. \textbf{108}, 023601 (2012).

\bibitem{Lahini1} Y. Lahini, M. Verbin, S. D. Huber, Y. Bromberg, R. Pugatch, and Y. Silberberg, Phys. Rev. A \textbf{86}, 011603(R) (2012).



\bibitem{Omar1} Y. Omar, N. Paunkovi\v{c}, L. Sheridan, and S. Bose, Phys. Rev. A \textbf{74}, 042304 (2006).

\bibitem{Benedetti1} C. Benedetti, F. Buscemi, and P. Bordone, Phys. Rev. A \textbf{85}, 042314 (2012).

\bibitem{Qin1} X. Qin, Y. Ke, X. Guan, Z. Li, N. Andrei, and C. Lee, Phys. Rev. A \textbf{90}, 062301 (2014).

\bibitem{Preiss1} P. M. Preiss, R. Ma, M. E. Tai, A. Lukin, M. Rispoli, P. Zupancic, Y. Lahini, R. Islam, and M. Greiner, Science \textbf{347}, 1229 (2015).

\bibitem{Wang2} L. Wang, L. Wang, and Y. Zhang, Phys. Rev. A \textbf{90}, 063618 (2014).

\bibitem{Lau:2020} L. L. H. Lau, and S. Dutta, Quantum walk of two anyons across a statistical boundary, arXiv:2012.03977.

\bibitem{Ganahl1} M. Ganahl, E. Rabel, F. H. Essler, and H. G. Evertz, Phys. Rev. Lett. \textbf{108}, 077206 (2012).

\bibitem{Krapivsky1} P. L. Krapivsky, J. M. Luck, and K. Mallick, J. Phys. A \textbf{48}, 475301 (2015).

\bibitem{Wiater1} D. Wiater, T. Sowi\'{n}ski, and J. Zakrzewski, Phys. Rev. A \textbf{96},043629 (2017).

\bibitem{Iyer:2013}D. Iyer, H.  Guan, and N.  Andrei,  Phys. Rev. A \textbf{87}, 053628 (2013).


\bibitem{Siloi:2017}I. Siloi, C. Benedetti, E. Piccinini, J. Piilo, S. Maniscalco, M. G. A. Paris, and P. Bordone, Phys. Rev. A {95}, 022106 (2017).

\bibitem{Beggi:2018}A. Beggi, I. Siloi, C. Benedetti, E. Piccinini, L. Razzoli, P. Bordone and M. G. A. Paris, Eur. J. Phys. {39}, 065401 (2018)

\bibitem{Sarkar:2020}S. Sarkar and T. Sowi\'{n}ski,  Phys. Rev. A \textbf{102}, 043326 (2020).

\bibitem{Liu1} W. Liu and N. Andrei, Phys. Rev. Lett. \textbf{112}, 257204 (2014).

\bibitem{Geiger1} Z. A. Geiger, K. M. Fujiwara, K. Singh, R. Senaratne, S. V. Rajagopal, M. Lipatov, T. Shimasaki, R. Driben, V. V. Konotop, T. Meier and D. M. Weld, Phys. Rev. Lett. \textbf{120}, 213201 (2018).

\bibitem{Ferrari1} G. Ferrari, N. Poli, F. Sorrentino, and G. M. Tino, Phys. Rev. Lett. \textbf{97}, 060402 (2006).

\bibitem{Tarallo1}    M. G. Tarallo, T. Mazzoni, N. Poli, D. V. Sutyrin, X. Zhang, and G. M. Tino, Phys. Rev. Lett. \textbf{113}, 023005 (2014).

\bibitem{LiuWJ} W. Liu, Y. Ke, L. Zhang, and C. Lee, Phys. Rev. A \textbf{99}, 063614 (2019).

\bibitem{Atala1} M. Atala, M. Aidelsburger, J. T. Barreiro, D. Abanin, T. Kitagawa, E. Demler and I. Bloch, Nat. Phys. \textbf{9}, 795(2013).

\bibitem{Carusotto1} I. Carusotto, L. Pitaevskii, S. Stringari, G. Modugno, and M. Inguscio, Phys. Rev. Lett. \textbf{95}, 093202(2005).



\bibitem{Helstrom:1976}C. W. Helstrom, Quantum Detection and Estimation Theory, Academic Press, New York, 1976.

\bibitem{Pezze1} L. Pezz\`{e}, A. Smerzi, M. K. Oberthaler, R. Schmied, and P. Treutlein, Rev. Mod. Phys. \textbf{90}, 035005 (2018).

\bibitem{Liu2} J. Liu, H. Yuan, X. M. Lu and X. Wang, J. Phys. A, \textbf{53}, 023001 (2020).


\bibitem{Sun:2020e} H. Sun, B. Yang, H. Y. Wang, Z. Y. Zhou, G. X. Su, H. N. Dai, Z. S. Yuan, J. W. Pan, arXiv: 2009.01426.

\bibitem{Yang:2020e}B. Yang, H. Sun, R. Ott, H. Y. Wang, T. V. Zache, J. C. Halimeh, Z. S. Yuan, P. Hauke, J. W. Pan, arXiv:2003.08945.


\bibitem{Jordan1} P. Jordan and E.Wigner, Z. Phys. \textbf{47}, 631 (1928).


\bibitem{YY:1966} C. N. Yang,  and C. P.  Yang,  Phys. Rev. \textbf{150}, 321 (1966);  \textbf{150},  327 (1966); \textbf{151}, 258 (1966).

\bibitem{Takahashi1} M. Takahashi, \textit{Thermodynamics of One-Dimensional Solvable Models} (Cambridge University Press, Cambridge, 1999).

\bibitem{Lee2004} C. Lee, Phys. Rev. Lett. \textbf{93}, 120406 (2004).

\bibitem{Vidal1} G. Vidal, Phys. Rev. Lett. \textbf{91}, 147902 (2003); Phys. Rev. Lett. \textbf{93}, 040502 (2004).


\bibitem{Yu1} J. Yu, N. Sun, and H. Zhai, Phys. Rev. Lett. \textbf{119}, 225302 (2017).


\bibitem{Supplemental} In the Supplemental Material, we give more details for the symmetries in dynamic evolution, correlation functions in three-particle QWs, the derivation of effective single-particle models for co-walkings and co-BOs, the two-body physics in three-particle QWs, the calculation of Fisher information and additional figures for spectra, QWs and BOs.

\bibitem{Takahashi1} M. Takahashi, J. Phys. C, \textbf{10}, 1289 (1977).


\bibitem{Kosevich1} Y. A. Kosevich, and V. V. Gann, J. Phys. Condens. Matter \textbf{25}, 246002(2013).

\bibitem{Zhang1} H. Zhang, Y. Zhai, and X. Chen, J. Phys. B \textbf{47}, 025301(2014).

\bibitem{Dias1} W. S. Dias, E. M. Nascimento, M. L. Lyra, and F. A. B. F. de Moura, Phys. Rev. B \textbf{76}, 155124 (2007).

\bibitem{Khomeriki1} R. Khomeriki, O. Krimer, M. Haque, and S. Flach, Phys. Rev. A \textbf{81}, 065601 (2010).

\bibitem{Corrielli1} G. Corrielli, A. Crespi, G. Della Valle, S. Longhi, and R. Osellame. Nat. Commun. \textbf{4}, 1555 (2013).

%
%
%
%
%
%


%
\bibitem{Breid1} B. M. Breid, D. Witthaut, and H. J. Korsch, New J. Phys. \textbf{8}, 110 (2006).
%
\bibitem{Longhi1} S. Longhi, Phys. Rev. B \textbf{86}, 075144 (2012).

%



\bibitem{Braunstein1} S. L. Braunstein and C. M. Caves, Phys. Rev. Lett. \textbf{72}, 3439 (1994).

\bibitem{Braunstein2} S. L. Braunstein, C. M. Caves and G. J. Milburn, Ann. Phys. \textbf{247}, 135 (1996).

\bibitem{footnote} The numerical values $\alpha =1185.485,\, 720.741\, 32.434,\, 3.604, \,0.024$ for the states 3B, S, 2B, 2F, 3F, respectively. The errors of such power-law simulation are less than 3\%.

\bibitem{footnote2} The third-order amplitude is proportional to $J^3/V^2$. The fourth and higher orders are also nontrivial but their amplitudes are much smaller than the third order when $J/V\ll1$.





\end{references}

\begin{references}
\bibitem{Yu1} J. Yu, N. Sun, and H. Zhai, Phys. Rev. Lett. \textbf{119}, 225302 (2017).

\bibitem{Qin1} X. Qin, Y. Ke, X. Guan, Z. Li, N. Andrei, and C. Lee, Phys. Rev. A \textbf{90}, 062301 (2014).

\bibitem{Takahashi1} M. Takahashi, J. Phys. C, \textbf{10}, 1289 (1977).

\bibitem{Liu1} J. Liu, H. Yuan, X. M. Lu and X. Wang, J. Phys. A, \textbf{53}, 023001 (2020).


\bibitem{Ferrari1} G. Ferrari, N. Poli, F. Sorrentino, and G. M. Tino, Phys. Rev. Lett. \textbf{97}, 060402 (2006).

\bibitem{Tarallo1}    M. G. Tarallo, T. Mazzoni, N. Poli, D. V. Sutyrin, X. Zhang, and G. M. Tino, Phys. Rev. Lett. \textbf{113}, 023005 (2014).


\end{references}
\end{document}